\documentclass[aps,prd,reprint,twocolumn,groupedaddress]{revtex4-1}

\usepackage{float}
\usepackage{algpseudocode}

\makeatletter
\newcounter{algorithm}
\renewcommand{\thealgorithm}{\arabic{algorithm}}
\newcommand{\fnum@algorithm}{Algorithm~\thealgorithm}

\newcommand{\ALG@caption@opt}[2][]{%
  \refstepcounter{algorithm}%
  \@fs@capt{\fnum@algorithm}{#2}\@fs@mid
}
\newcommand{\ALG@caption@noopt}[1]{\ALG@caption@opt[]{#1}}

\newenvironment{algorithm}[1][tbp]{%
  \begin{figure}[#1]%
    \fs@ruled
    \def\@captype{algorithm}%
    \@fs@pre
    \let\caption\relax
    \def\caption{\@ifnextchar[{\ALG@caption@opt}{\ALG@caption@noopt}}%
}{%
    \@fs@post
  \end{figure}%
}

\newenvironment{algorithm*}[1][tbp]{%
  \begin{figure*}[#1]%
    \fs@ruled
    \def\@captype{algorithm}%
    \@fs@pre
    \let\caption\relax
    \def\caption{\@ifnextchar[{\ALG@caption@opt}{\ALG@caption@noopt}}%
}{%
    \@fs@post
  \end{figure*}%
}
\makeatother

\usepackage{graphicx} 
\usepackage{dcolumn}  
\usepackage{bm}       
\usepackage{amsmath}  
\usepackage{amssymb}  
\usepackage{dsfont}
\usepackage{bbold}
\usepackage{nicematrix}
\usepackage[colorlinks=true, urlcolor=violet, linkcolor=blue, citecolor=red, hyperindex=true, linktocpage=true, draft=false]{hyperref}

\usepackage{tikz-cd}
\usepackage{xcolor}
\usepackage{multirow}
\usepackage{tablefootnote}
\usetikzlibrary{positioning}

\definecolor{Pank}{rgb}{1.0,0.1,0.5}

\usepackage{cancel}

\usepackage{thm-restate}

\newtheorem{lemX}{Lemma}

\definecolor{darkblue}{rgb}{0,0,.6}
\definecolor{orange}{rgb}{1,0.5,0}

\usepackage{comment}

\usepackage{wasysym,mdframed,tikz,leftindex,booktabs}

\usetikzlibrary{patterns,arrows}
\NiceMatrixOptions{cell-space-limits = 2pt}

\newcommand{\I}{\mathrm{I}} 
\newcommand{\J}{\mathrm{J}}
\newcommand{\T}{\mathrm{T}}

\newcommand{\F}{\mathrm{F}}
\newcommand{\M}{\mathrm{M}}

\newcommand{\ketzero}{|0\rangle}
\newcommand{\ketone}{|1\rangle}

\newcommand{\blue}[1]{\textcolor{blue}{#1}}

\definecolor{nicegreen}{rgb}{0.13, 0.55, 0.13}

\begin{document}

\title{Full grid solution for multi-asset options pricing with tensor networks}

\author{Lucas Arenstein}\email{lsa@di.ku.dk}
\author{Michael Kastoryano}

\affiliation{Department of Computer Science, University of Copenhagen, Denmark}

\date{\today}

\begin{abstract}
Pricing multi-asset options via the Black--Scholes PDE is limited by the curse of dimensionality: classical full–grid solvers scale exponentially in the number of underlyings and are effectively restricted to three assets. Practitioners typically rely on Monte Carlo methods for computing complex instrument involving multiple correlated underlyings. We show that quantized tensor trains (QTT) turn the $d$-asset Black--Scholes PDE into a tractable high-dimensional problem on a personal computer. We construct QTT representations of the operator, payoffs, and boundary conditions with ranks that scale polynomially in $d$ and  polylogarithmically in the grid size, and build two solvers: a time-stepping algorithm for European and American options and a space--time algorithm for European options. We compute full-grid prices and Greeks for correlated basket and max--min options in three to five dimensions with high accuracy. The methods introduced can comfortably be pushed to full-grid solutions on $10-15$ underlyings, with further algorithmic optimization and more compute power. 
\end{abstract}

\maketitle

\section{Introduction}
Pricing multi-asset options remains a central challenge in quantitative finance. 
In the Black--Scholes model with $d$ correlated underlyings, the natural deterministic route is to solve a parabolic partial differential equation (PDE) in $d$ spatial dimensions.
However, classical grid–based methods (finite differences, finite elements or spectral methods) require a grid whose size grows in 
${\mathcal{O}(N^d)}$ for $N$ discretization points per dimension causing both runtime and memory to explode exponentially. This phenomenon, commonly referred to as the \emph{curse of dimensionality}, effectively limits these classical methods to three-asset problems.

By abandoning the full grid, \emph{sparse-grid} techniques~\cite{bungartz2004sparse} alleviate the curse of dimensionality, reducing the number of grid points from $\mathcal{O}(N^d)$ to $\mathcal{O}(N (\log N)^{d-1})$, and have been successfully applied to Black--Scholes PDEs for $d>3$~\cite{heinecke2012option,pfuger2010adaptive}. However, since these methods do not provide the solution on the full grid but in a hierarchical representation, tasks such as computing Greeks on a fixed lattice, re-pricing when the underlying asset values move slightly away from the original grid points, and the overall implementation complexity all become significant bottlenecks. 

As a result, practitioners often resort to Monte Carlo (MC) and  least-squares algorithms \cite{longstaff2001valuing} for early–exercise features, accepting stochastic noise, slower convergence for tail events, and---crucially---the lack of a full solution surface: MC gives prices (and noisy Greeks) at specific initial states and must be rerun for every new scenario.

A full state–space solution enables practical capabilities that sparse-grid techniques and MC based pipelines cannot match. In this work we demonstrate:
(i) \emph{instant re-pricing} for new spots within the domain via interpolation, without reruns; and
(ii) \emph{dense Greeks}—including cross-sensitivities—computed consistently over the grid.
We further note that the same full-grid representation naturally supports (though we do not empirically demonstrate them here):
(iii) \emph{calibration} to entire surfaces rather than isolated points, which can improve stability in inverse problems (implied volatility or correlation).

These capabilities are directly relevant to hedging, intraday risk, and model validation. In practice, the solution grid can be (re)built during market hours on commodity hardware; thereafter, (re)pricing and Greek queries are essentially instantaneous via fast grid interpolations. This stands in sharp contrast to workflows that typically rely on overnight Monte Carlo runs.

This paper advocates a quantized tensor train (QTT) approach which allows to represent and manipulate high dimensional functions and operators efficiently. 
By encoding all the elements of the PDE in QTT form, we convert exponential–in–$d$ storage and arithmetic into computations that scale polylogarithmic in $N$.

We develop two complementary solvers:
(i) a \emph{time–stepping QTT} algorithm for European and American options (with early–exercise enforced directly in QTT form), and
(ii) a \emph{space–time QTT} algorithm for European options that treats time as an extra dimension and computes the entire temporal evolution in one shot. 
To our knowledge, this is the first work that delivers the full-grid solutions of the multi-asset Black--Scholes PDE beyond the classical three-asset barrier on a personal computer.

\paragraph{Related work.}
For high-dimensional elliptic and parabolic PDEs, quantized tensor trains have emerged as state-of-the-art among deterministic solvers, offering polylogarithmic-in-grid complexity~\cite{qtt_khoromskij,richter2021solving}. 
A domain that particularly benefits from such compression is computational fluid dynamics (CFD), where extremely large grids are required: recent tensor-network (TN) pipelines report compressions and end-to-end computations that are impractical with classical dense/sparse representations~\cite{gourianov2022exploiting,peddinti2024quantum,pisoni2025compression,van2025quantum,siegl2025tensor,amaral2025quantum}.

Within space–time formulations, several variants exist. 
DMRG-inspired solvers for linear and nonlinear PDEs were presented in \cite{dolgov2012fast, arenstein2025fast, peddinti2025quantum}. For Riccati-type problems, space–time tensor methods have been combined with Newton–Kleinman iterations~\cite{breiten2019solvingdifferentialriccatiequations}. 
Spectral space–time collocation has been realized in tensor network (TN) form~\cite{adak2024tensornetworkspacetimespectral,adak2024space}, and finite-element space–time discretizations have likewise been adapted to TN frameworks~\cite{adak2025space}. 

In this work, we adopt the time-stepping and space--time framework of~\cite{arenstein2025fast}, which exhibits low time complexity under modest TT-rank assumptions—consistent with our benchmarks—and generalizes naturally to high-dimensional PDEs.

Within options pricing, tensor networks were first deployed to accelerate multi-asset Fourier pricing in~\cite{kastoryano2022fourierTN}. Building on this Fourier framework a tensor train learning algorithm to price options was proposed in \cite{sakurai2024ttparam} and a TT formulation for efficient computation of Greeks~\cite{sakurai2025ttgreeks}.

Beyond Fourier methods, low-rank tensor approximations based on Chebyshev interpolation have been used for option pricing~\cite{glau2019lowranktensorapproximationchebyshev}; lattice/tree schemes have been boosted via MPS/TT representations~\cite{vandamme2025tnbinomial}; tensor-structured neural networks have been proposed for pricing~\cite{patel2022tnn}; and TNs have been leveraged for path generation in stochastic models relevant to valuation~\cite{kobayashi2024tnpaths}.

In contrast to these lines, our work takes a fundamentally different route: we present the first tensor-network (QTT) solvers to compute full-grid solutions of the multi-asset Black--Scholes PDE for pricing (and Greeks) -- running on a laptop.

\paragraph{Contributions.}
\begin{itemize}
  \item \textbf{Two solvers.}
        (a) A full-grid \emph{time–stepping QTT} algorithm for European and American multi-asset options, with Greeks computed consistently on the grid.
        (b) A \emph{space–time QTT} algorithm for European multi-asset options that solves all time layers simultaneously and returns the full temporal evolution in one run and likewise supports dense Greeks calculation.
    \item \textbf{QTT formulation of the $d$-asset Black--Scholes operator and problem data.}
        We construct the Black--Scholes operator exactly in QTT with provable rank bounds that grow polynomially in $d$ (and not with the grid size). We further show that the payoff, initial and boundary conditions admit low TT ranks at the target accuracy, making iterative solves practical in high dimensions.

    \item \textbf{Full-grid $d>3$ correlated options.}
    We obtain the first high accuracy, full grid solution for a correlated option pricing problem in dimensions larger than three.
\end{itemize}

\paragraph{Main findings.}
(i) In $d=3,4,5$ we obtain accurate \emph{full grids} on a laptop, with run times reported against high–accuracy references. 
(ii) For European options, space--time is often faster for $d\le 3$ and uniquely gives the full temporal surface; for $d>3$, time--stepping is typically preferable.
(iii) For Americans, the early–exercise projection adds only a modest overhead; keeping QTT ranks controlled is the dominant factor, not the raw grid size.

\paragraph{Paper organization.}
Section~\ref{sec:multi_asset_option} recalls single-asset contracts and introduces the multi-asset setting, culminating in the $d$-dimensional Black--Scholes PDE. 
Section~\ref{sec:basics} presents our tensor-network methodology with the main elements to build the solvers.
Section~\ref{sec:space_time_main} introduces the space–time QTT formulation for Black--Scholes, contrasting it with time–stepping.
Section~\ref{sec:numerics_euro} reports numerical experiments for European options; Section~\ref{sec:numerics_american} extends the study to American and analyzes the extra cost of enforcing the early-exercise condition. In the Appendices we provide all algorithms and time-complexity derivations used in this paper and selected supplementary experiments.

\section{Multi-asset Options}\label{sec:multi_asset_option}

Before moving to the multi-asset setting, we briefly recall the single-asset case.  A \emph{European put option} on a single underlying asset grants the holder (buyer of the contract) the right, but not the obligation, to sell one unit of that asset to the option writer (seller of the contract) for a fixed strike price $K$ on a predetermined maturity date $T$. The underlying may be a stock, bond, foreign-exchange rate, commodity, or any other instruments.  At expiry the contract value (payoff) is $\max(K-S_T,0)$, where $S_T$ is the asset price at time $T$.

If the holder is allowed to exercise at any time ${t\in[0,T]}$, the contract is called an \emph{American put option}.  The early-exercise feature can only add value, so an American put is never worth less than its European counterpart with the same parameters. 

A \emph{European call option} grants the holder the opposite right, but not the obligation, to purchase the underlying asset for the strike price $K$ on the maturity date $T$. Its payoff is $\max(S_T-K,0)$. If the holder may exercise at any time $t\in[0,T]$, the contract becomes an \emph{American call option}. 

For European contracts the call and put prices are linked by the 
\emph{put–call parity}: $C_0 - P_0 = S_0 - K e^{-r T}$, where $C_0$ and $P_0$ denote the prices at time $0$ of the call and put with the
same strike $K$, maturity $T$ and $r$ is the risk-free interest rate. From a numerical standpoint this identity is useful in two ways. First, it provides an inexpensive consistency check: computing both prices with the same solver should reproduce the parity to within discretization error. Second, if one side is easier (e.g., the put in a dividend-free market), the call follows algebraically from the parity.

A \emph{multi-asset} put option extends the single-asset contract to a collection of underlying assets.
In the $d$-asset \emph{basket} put, the holder has the right to sell a specified linear combination of the assets typically the weighted sum
$B_t=\sum_{i=1}^{d}\omega_i S_t^{(i)}$, with weights $\omega_i>0$ for the strike price $K$. The payoff is therefore $\max(K-B_t,0)$.

Variants include the \emph{max-put} and \emph{min-put} contracts, in which the basket $B_t$ is replaced by the maximum or minimum component price, giving payoffs ${\max\bigl(K-\max(S_t^{(i)}),0\bigr)}$ and ${\max\bigl(K-\min(S_t^{(i)}),0\bigr)}$ with $1\leq i \leq d$, respectively. Many other pay-off structures are possible, for example, Haug’s book \cite[Chap 5]{haug2007complete} lists more than a dozen additional multi-asset option types. 

Except for a few special cases multi-asset pay-offs admit no closed-form valuation. In particular, all multi-asset American contracts and important Europeans contracts can only be priced numerically, motivating the high-dimensional methods developed in this work.

In both single and multi-asset setting the main challenge in option pricing is to determine each contract’s fair value at any time ${t\le T}$. We address this task for the multi-asset European and American put options framing the problem within the Black--Scholes partial differential equation (PDE) formalism. For a single underlying asset with price $S>0$, volatility $\sigma$, and constant risk-free interest rate $r$, the Black--Scholes PDE for the option price $V(S,t)$ at time $t \in [0,T)$ is:
\begin{align*}
    \frac{\partial V}{\partial t} + 
    \frac{1}{2}\sigma^2 S^2 \frac{\partial^2 V}{\partial S^2} + 
    r S \frac{\partial V}{\partial S}  
     - rV = 0.
\end{align*}
Extending this to $d$ underlying assets
$S=(S_1,\dots,S_d)^{\!\top}$, with volatilities $\sigma_i$ and positive definite correlation matrix $\rho = (\rho_{ij})$
of the underlying assets, we obtain
\begin{equation}\label{eq:multi_asset_BS}
\begin{aligned}
\frac{\partial V}{\partial t}
  + \frac{1}{2}\sum_{i=1}^{d}\sigma_i^{2} S_i^{2}
        \frac{\partial^{2}V}{\partial S_i^{2}}
  + \frac{1}{2}\sum_{\substack{i,j=1\\ i\ne j}}^{d}
        \rho_{ij}\sigma_i\sigma_j S_i S_j
        \frac{\partial^{2}V}{\partial S_i\,\partial S_j}\\
  + \sum_{i=1}^{d} r S_i\frac{\partial V}{\partial S_i}
  - rV = 0.
\end{aligned}
\end{equation}
This $d$-dimensional Black--Scholes equation is the starting point for the tensor-based solvers developed in the remainder of the paper.

\section{Solving the Black--Scholes with Tensor Networks}\label{sec:basics}

This section introduces the building blocks of our QTT-based solver for the $d$-dimensional Black--Scholes equation \eqref{eq:multi_asset_BS}.
We begin by discretizing the PDE with an implicit finite-difference scheme, turning each time step into a linear system $Aw^{i+1}=b^i$.
With our chosen discretization scheme the $A$ matrix will have a structured form that we will explore to build an exact low-rank quantized tensor train (QTT) representation of this operator. 
We then compress the \textit{right-hand side} (RHS) $b^i$ which contains the terminal pay-off (for $i=0$) and time-dependent boundary conditions (for $i>0$) using analytic QTT constructions and the \emph{TT-Cross} \cite{oseledets2010ttcross} algorithm, obtaining a matching low-rank representation.
Finally, the resulting system is entirely solved in the QTT framework with the Alternating Linear Scheme (ALS)\cite{als_main}.

\subsection{Tensor train basics}\label{sec:tt}

We briefly review the tensor network terminology used in this work.
A tensor $\T\in\mathbb{R}^{n_1\times\cdots\times n_d}$ is a {$d$-dimensional} array (also called an \emph{order-$d$} tensor), its
entries are written $\T_{x_1,\dots,x_d}$ with
$1\le x_i\le n_i$. Each $n_i$ is called a \emph{mode}. Large tensors can be decomposed into a network of smaller ones by summing over shared indices. A prominent type of tensor network is the tensor train \cite{tt_ose} (TT) format. A tensor is in TT form if it factorizes as
\begin{align*}
T_{x_1,x_2,...,x_d}=T_1^{x_1} T_2^{x_2} \cdots T_d^{x_d},
\end{align*}
where $T_j \in \mathbb{R}^{R_{j-1}\times n_j \times R_j}$ are \emph{TT cores}. The integers $R_j$ are called \emph{bond dimensions} or \emph{TT ranks} and measure the correlation carried from one core to the next. In the TT above, we note that $R_0 = R_d = 1$ and $R_j \geq 1$ for $2 \leq j \leq d-1$.
We will generally refer to \textit{the} TT-rank as the maximum of all TT ranks in the given tensor train. 

Graphical notation offers an intuitive way to visualize tensor networks. For example, a TT with four cores is depicted as:
\tikz[baseline=-0.5ex]{
    \node[draw, circle, inner sep=1pt] (tensor1) {\scriptsize $T_1$};
    \node[draw, circle, inner sep=1pt, right=0.3cm of tensor1] (tensor2) {\scriptsize $T_2$};
    \node[draw, circle, inner sep=1pt, right=0.3cm of tensor2] (tensor3) {\scriptsize $T_3$};
    \node[draw, circle, inner sep=1pt, right=0.3cm of tensor3] (tensor4) {\scriptsize $T_4$};

    \draw (tensor1) -- (tensor2);
    \draw (tensor2) -- (tensor3);
    \draw (tensor3) -- (tensor4);

    \draw (tensor1) -- ++(0,-0.4);
    \draw (tensor2) -- ++(0,-0.4);
    \draw (tensor3) -- ++(0,-0.4);
    \draw (tensor4) -- ++(0,-0.4);
}, circles represent the cores $T_1,\dots,T_4$, horizontal lines are the bond dimensions ($R_j$), and the downward legs denote the \emph{physical dimensions} (indices of size $n_j$).

An important extension is the \emph{TT-operator} or MPO, which
decomposes a tensor into a chain of order-4 cores of size $(R_{j-1},n_j,m_j,R_j)$. The free indices $n_j$ and $m_j$ are the physical dimensions and the contracted indices $R_j$ are the bond dimensions. Graphically a $4$ cores TT operator is represented in a similar fashion to the TT above: 
\tikz[baseline=-0.5ex]{
    \node[draw, circle, inner sep=1pt] (tensor1) {\scriptsize $T_1$};
    \node[draw, circle, inner sep=1pt, right=0.3cm of tensor1] (tensor2) {\scriptsize $T_2$};
    \node[draw, circle, inner sep=1pt, right=0.3cm of tensor2] (tensor3) {\scriptsize $T_3$};
    \node[draw, circle, inner sep=1pt, right=0.3cm of tensor3] (tensor4) {\scriptsize $T_4$};

    \draw (tensor1) -- (tensor2);
    \draw (tensor2) -- (tensor3);
    \draw (tensor3) -- (tensor4);

    \draw (tensor1) -- ++(0,-0.4);
    \draw (tensor2) -- ++(0,-0.4);
    \draw (tensor3) -- ++(0,-0.4);
    \draw (tensor4) -- ++(0,-0.4);

    \draw (tensor1) -- ++(0,0.4);
    \draw (tensor2) -- ++(0,0.4);
    \draw (tensor3) -- ++(0,0.4);
    \draw (tensor4) -- ++(0,0.4);
} but with $4$ extra upward legs to denote the indices $m_j$.

One of the main advantages of working on the TT representation of vectors and operators is that they provide compact storage and allow key linear operations such as matrix vector products to be carried out locally, yielding dramatic savings in memory and runtime.

\subsection{The quantized tensor train}\label{sec:qtt}

The QTT format compresses regular grids by reshaping long modes into a
sequence of binary ones.  Consider a scalar function
$f\colon[0,1)\to\mathbb{R}$ sampled on the uniform grid
$\{k/2^{c}\}_{k=0}^{2^{c}-1}$, for integer $c$.  Writing
$k=(x_1x_2\ldots x_{c})_{2}$ in binary identifies each sample with the
bit–tuple ${(x_1,\dots,x_{c})\in\{0,1\}^{c}}$, so the data can be viewed as the
tensor
$$
f_{x_1x_2\ldots x_{c}}
   = f\!\bigl(0.x_1x_2\ldots x_{c}\bigr).
$$
The single mode of length $2^{c}$ is thus replaced by $c$ modes of length~2.
We approximate this $2\times\cdots\times2$ tensor by a special case of a $c$-cores TT decomposition that is called the \emph{quantized tensor train} (QTT): 
$$
f_{x_1\ldots x_{c}}
   = Q^{x_1}_1\,Q^{x_2}_2\cdots Q^{x_{c}}_{c},
$$
where $Q^{x_j}_j\in\mathbb{R}^{R_{j-1}\times 2 \times R_{j}}$ and $R_0=R_{c}=1$.

If all ranks (bond dimension) satisfy $R_j\le \chi$, the storage falls from $\mathcal{O}(2^{c})$ to $\mathcal{O}(c\,\chi^{2})$, i.e.\ logarithmic in the grid size.  Because bit positions correspond to dyadic length scales, the cores capture multiresolution structure. As a result, many piecewise-smooth functions, such as sine or exponential (see Appendix \ref{appen:analytic_qtt_exp}), admit very small QTT ranks \cite{kazeev2012}.The idea extends dimension-wise: a $d$-dimensional tensor on a $2^{c}\times\!\cdots\times2^{c}$ grid is reshaped into $dc$ binary modes and stored as a QTT at cost $\mathcal{O}(d\,c\,\chi^{2})$.

Just as a tensor train generalizes to a TT-operator (MPO), we can also build a \emph{QTT-operator} by adding, for each core, a second physical leg of size 2. This yields a chain of order-4 binary cores
$(R_{j-1},2,2,R_j)$ that can act on a QTT representation of a vector.
An important example is the QTT-operator representation of the tridiagonal-constant (Toeplitz) matrix. It was proven in \cite{kazeev2012} that there exists an exact rank-3 QTT representation of this matrix. We have restated this lemma in Appendix \ref{appen:lemma_1} with our notation. In our Black--Scholes QTT solver we choose finite-difference stencils so that each differential operator factors into a Kronecker product of such matrices.

\subsection{TT-Cross}\label{sec:tt_cross}

Consider a multivariate function
${f:[0,1]^{d}\rightarrow\mathbb{R}}$ discretized into $2^c$ points per dimension that can be evaluated in any one of the dyadic grid points
$\{0,1/2^{c},\ldots,(2^{c}-1)/2^{c}\}^{d}$.
If our goal is to build the QTT representation of this function we first note that sampling all the $2^{c d}$ points is infeasible when $d$ or $c$ is moderate.

The \emph{TT-Cross} algorithm \cite{oseledets2010ttcross} circumvents
this bottleneck by sampling only $\mathcal{O}(d\,c\,\chi^{2})$ carefully chosen points and building a QTT approximation of $f$ whose TT ranks do not exceed a user-prescribed $\chi$.

In our solver, TT-Cross is used primarily to implement the non-linear
$\max$ operation on QTT objects. If an initial or boundary condition admits a closed-form QTT representation, for instance, an exponential we first construct that representation and then apply TT-Cross to compute the $\max$ between this representation and a scalar or a function. For American options, the same procedure enforces the early-exercise condition at each time step by evaluating $\max\{\text{current option value},\text{payoff}\}$ directly in the QTT format.

\subsection{Alternating Linear Scheme}\label{sec:als}

Once we have the QTT representation of the operator $A$ and the RHS $b$ we are ready to solve the QTT representation of the linear system $Aw=b$. For this we use the \emph{Alternating Linear Scheme} (ALS) \cite{als_main} whose output is the QTT representation of the solution $w$.

ALS solves a linear system in TT/QTT format by optimizing one core at a time while keeping the others fixed, whereas its MALS variant optimizes a pair of cores simultaneously. At every step all contracted cores except the active one are absorbed into two “environment” tensors, turning the global equation $Aw=b$ into a much smaller dense system whose unknowns are the entries of the active core. This local system is solved, and the algorithm moves to the next core, sweeping left-to-right and right-to-left.

For a QTT with $c$ cores per $d$ dimension the cost of one full sweep is
$\mathcal{O}(4c d \gamma \chi_A^2 \chi^3)$, where
$\chi_A$ and $\chi$ are the maximal ranks of $A$ and $w$ (or $b$),
respectively, and $\gamma$ is the iteration count of the local solver. We note that this time complexity is fundamentally different from the best classical solvers whose complexity is $\mathcal{O}(2^{cd})$. On the QTT framework we are not limited by the number of discretization points but our challenge is to keep the elements in the low-rank regime. In Section \ref{sec:numerics}, we show that our overall complexity remains polynomial through numerical experiments and later present a time complexity analysis (Section \ref{app:d-asset_alg}) of the whole algorithm that agrees with these experiments. These results confirm our exponential speedup over classical grid based methods.

\subsection{QTT Time-Stepping Algorithm Overview}\label{sec:qtt_ts_alg}

\begin{figure*}[!t]  
    \centering
    \includegraphics[width=\textwidth,height=\textheight,keepaspectratio]{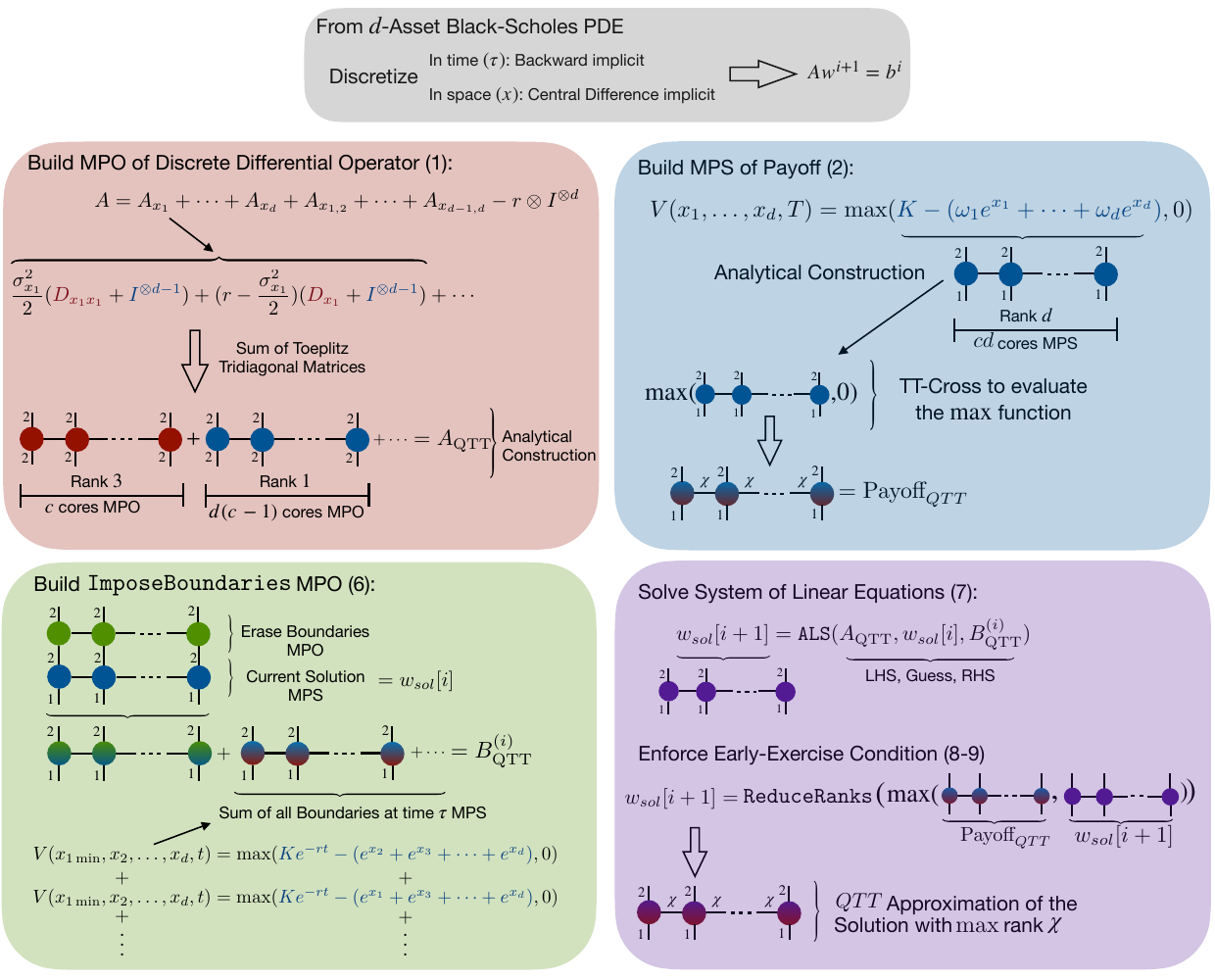}
    \caption{Schematic overview of the QTT time-stepping pipeline for the $d$-asset Black--Scholes PDE. 
    The grey box starts from Eq.~\eqref{eq:multi_asset_BS}, applies a backward Euler discretization in time and centered finite differences in the log–prices, and yields a linear system at each time level that is never formed explicitly.  The red box builds a matrix-product-operator (MPO) representation of the
    discrete differential operator.  The blue box constructs the QTT representation of the payoff. Using an analytical construction for the ``blue terms'' and TT-Cross to evaluate the $\max$ function.
    The green box imposes the boundary conditions at each time step: each boundary is encoded in QTT form using the same procedure as for the payoff.  Finally, the purple box solves the resulting linear system in the QTT framework and, in the American case, applies the early-exercise condition and a rank reduction of the solution (lines~8--9 of Alg.~\ref{alg:qtt_ts_main}).}
    \label{fig:main}
\end{figure*}

Combining the building blocks above, we can build the QTT solvers used throughout the paper. Figure~\ref{fig:main} shows the full pipeline of the QTT time-stepping solver for the American $d$-asset Black--Scholes PDE, highlighting the key tensor-network operations. The mathematical derivation and QTT constructions are given in Appendices~\ref{app:2-asset} and~\ref{app:d-asset_alg}. The complete pseudocode is provided in Sec.~\ref{sec:numerics_american}, the line numbers there correspond to the steps annotated in Fig.~\ref{fig:main}.

\section{Space–Time QTT Solver for the Black--Scholes Equation}\label{sec:space_time_main}

Classical grid-based methods for time-dependent PDEs typically treat time and space as separate variables. 
In a time-stepping scheme, the PDE is discretized in space, and the solution is advanced sequentially through time: at each time level the system is solved for the spatial variables, and the resulting state is used to compute the next one.
An alternative approach is to treat time as an additional spatial dimension and solve the full space–time system simultaneously. 
However, this formulation quickly becomes computationally prohibitive for classical solvers. Even a three-dimensional spatial problem with one temporal dimension corresponds to a four-dimensional PDE, which exceeds the limits of conventional grid-based methods.

In \cite{arenstein2025fast}, the authors demonstrated that the QTT space-time approach is feasible, and can outperform time-stepping in certain situations. We demonstrated this property on the one-dimensional Burgers’ equation. Along the same lines, Peddinti et al.~\cite{peddinti2025quantum} showed that the same space--time formulation can be used to solve a broader class of linear and non-linear PDEs.

In Appendix~\ref{appen:space_time_1D}, we extend this QTT space–time formulation to the one-asset Black--Scholes equation and, in Appendix~\ref{appen:space_time_analysis}, generalize the construction to the $d$-asset case, deriving an upper bound on the algorithm’s time complexity. In Appendix ~\ref{app:space_time_greeks} we show how we can efficiently compute the Greeks in this framework.

Specifically, we solve the Black–Scholes equation using a space–time discretization that treats time as an additional spatial dimension, producing a structured linear system well suited for low-rank QTT compression. After transforming variables and discretizing the PDE on a uniform grid, the resulting block–bidiagonal operator and the right-hand side—including the terminal payoff and both spatial boundary conditions—are shown to admit exact, low-rank QTT representations. This allows the entire space–time solution surface to be computed at once using QTT-based ALS solvers rather than via time-stepping. Boundary conditions are incorporated directly into the QTT representation of the right-hand side, ensuring they propagate correctly through the tensor solver. The Greeks are evaluated by finite difference on the lattice.



To highlight the difference between time–stepping and space–time approaches, it is useful to compare their solution representations.
In classical time–stepping, the solver produces a sequence of spatial solution vectors $w_t$, one for each time level, and the accuracy of the final result depends on the resolution of each individual time step.
In the space–time formulation, by contrast, the solver returns a single global vector that contains all time layers at once, so the entire solution surface—intermediate states and terminal values—is computed in one shot rather than propagated sequentially through time. Because QTT representations scale only logarithmically with the grid size, treating time and space on equal footing can be highly advantageous: if the bond dimensions remain moderate, the full space–time problem can be solved at a cost that grows polylogarithmically in the number of grid points, offering substantial compression relative to classical time-stepping.

For many PDEs of practical interest in engineering, access to the full temporal evolution of the solution is essential. 
However, in option pricing we are typically concerned only with the final time slice, which corresponds to the present value of the contract.
Under this restriction, the space–time formulation may appear less economical than the time–stepping one, since it computes the entire evolution surface rather than only its terminal state.
Nonetheless, for a moderate number of dimensions, the space–time method remains advantageous due to its favorable QTT scaling, and also because it circumvents the CFL condition, as discussed in Ref. \cite{arenstein2025fast}.

In Appendices~\ref{app:d-asset_alg} and~\ref{appen:space_time_analysis}, we derive upper bounds for the time complexity of both approaches. We consider the cost of producing a final solution of size $2^{cd}$, where $c$ is the number of cores per spatial dimension and $d$ the number of assets. Table~\ref{tab:ts_st_complexity} 
\begin{table}[h]
    \centering
    \begin{tabular*}{0.47\textwidth}{@{\extracolsep{\fill}}ccccc@{}}
        \toprule
        \textbf{Method} &
        \textbf{Dominant Scaling}\\
        \hline
        \midrule
        Time-stepping   & $\mathcal{O}\!\bigl(2^c\, (d^5 + d^3)\, \chi_{\text{TS}}^{3} \bigr)$ \\  
        Space--time  & ${\mathcal{O}\!\bigl( c(d^{5}+d^3)\,\chi_{\text{ST}}^{3}\bigr)}$ \\  
        \hline
        \bottomrule
    \end{tabular*}
    \caption{}\label{tab:ts_st_complexity}
\end{table}
reports the dominant asymptotic time complexity of the QTT time-stepping and space--time solvers for a $d$-asset Black--Scholes PDE on a grid of size $2^{cd}$, where $c$ is the number of QTT cores per spatial dimension and $d$ the number of assets. The factor $2^{c}$ in the time-stepping method corresponds to the number of time steps.
The common multiplicative constant in both bounds is the square of the maximum bond dimension of the discrete differential operator $A$.

The quantities $\chi_{\mathrm{TS}}$ and $\chi_{\mathrm{ST}}$ denote the maximal QTT ranks of the RHS tensors in the two formulations.
For time-stepping, $\chi_{\mathrm{TS}} := \max \mathrm{rank}(B_\text{QTT}^{(i)})$, where $B_\text{QTT}^{(i)}$ is the QTT representation of the RHS of the system $A w^{i+1}=b^{i}$ after imposing the boundary conditions at time step $i$.
For space--time, $\chi_{\mathrm{ST}} := \mathrm{rank}(B_\text{QTT})$, where $Aw=b$ is the global space--time linear system and $B_\text{QTT}$ stacks the terminal payoff (initial condition) and all spatial boundary values across all time layers (see Eq.~\eqref{eq:b_vec} for the 1D case).

In practice, the time-stepping
solver employs a few initial MALS sweeps followed by ALS, whereas the space--time solver typically requires the two-cores optimization MALS to achieve the same accuracy.

Heuristically, if $2^{c}\,\chi_{\mathrm{TS}}^{3}\gtrsim\chi_{\mathrm{ST}}^{3}$ (ignoring constant factors), space–time wins; otherwise time-stepping tends to be faster. Time-stepping benefits by smaller ranks per step because the right hand side of this system encodes only the the boundary conditions that are enforced at each time step. On the contrary, the right hand side of the space--time solver has a more complex structure because it needs to enforce the boundary and initial condition on all the differential elements of the PDE, so without any compression we expect $\chi_{\text{ST}} >\chi_{\text{TS}}$.

In both formulations, the only substantial source of rank growth is the nonlinear $\max$ operator. Using TT-Cross to approximate the $\max$ term, our experiments indicate that space–time is often faster for $d \leq 3$, while for higher dimensions time–stepping becomes preferable.

Finally, for context, the best classical grid-based solvers scale at least as $\mathcal{O}(2^c\,2^{cd})$ in time, which is prohibitive beyond $d>3$, whereas the QTT approaches above remain polynomial in $d$.

\subsection{Benchmarks: Space–Time vs.\ Time–Stepping}

To compare the space--time and time-stepping solvers, we consider a one-asset European call option with strike $K = 65$, risk-free rate $r = 0.08$, volatility $\sigma = 0.3$, and time to maturity $T = 0.25$. 
Both solvers were tuned to reach the approximately the same mean absolute error over the full grid as quickly as possible, benchmarked against the known closed-form analytical solution. We also report the maximum absolute error observed at any grid point. All experiments were performed using the MALS algorithm with two sweeps and a spatial domain $[K/3, 3K]$, chosen to ensure the payoff and boundary conditions are accurately captured within the computational range.

\begin{table}[h]
    \centering
    \begin{tabular*}{0.47\textwidth}{@{\extracolsep{\fill}}ccccc@{}}
        \toprule
        \textbf{\shortstack{{Total}\\ {Cores}}} &
        \textbf{\shortstack{{Time}\\ {Steps}}} &
        \textbf{\shortstack{{Grid}\\ {Mean $|$Error$|$}}} &
        \textbf{\shortstack{{Max}\\ {$|$Error$|$ Grid}}} &
        \textbf{\shortstack{{Run}\\ {Time(s)}}}\\
        \hline
        \midrule
        6   & $2^5$   & 4.1e-03 & 5.0e-02  & 0.041\\  
        7  & $2^6$   & 1.6e-03  & 2.7e-02 & 0.119\\  
        8  & $2^7$ & 4.2e-04  & 4.9e-03 & 0.361\\  
        9  & $2^8$ & 2.0e-04  & 3.2e-03 & 1.071\\  
        \hline
        \bottomrule
    \end{tabular*}
    \caption{Time–stepping - MALS, two sweeps. The time step size is $1/(\# \text{time steps})$.}    \label{tab:eur_timestepping}
\end{table}

\begin{table}[h]
    \centering
    \begin{tabular*}{0.47\textwidth}{@{\extracolsep{\fill}}cccc@{}}
        \toprule
        \textbf{\shortstack{{Total}\\ {Cores}}} &
        \textbf{\shortstack{{Grid}\\ {Mean $|$Error$|$}}} &
        \textbf{\shortstack{{Max}\\ {$|$Error$|$ Grid}}} &
        \textbf{\shortstack{{Run}\\ {Time(s)}}}\\
        \hline
        \midrule
        10  & 7.1e-03  & 3.7e-02  & 0.01\\  
        12  & 2.3e-03  & 1.2e-02 & 0.02\\  
        14  & 8.8e-04  & 4.1e-03 & 0.07\\  
        16  & 4.0e-04  & 1.4e-03 & 0.14\\  
        \hline
        \bottomrule
    \end{tabular*}
    \caption{Space–time - MALS, two sweeps. ``Total Cores'' counts space + time QTT cores that are the same for this experiment.}
    \label{tab:eur_spacetime}
\end{table}

As expected in this low-dimensional setting, the space–time solver outperforms time-stepping. Examining the errors over the grid shows that both approaches can compute the option price for different initial values of the stock within the spatial domain. The space–time formulation additionally yields the full temporal evolution, whereas time-stepping primarily targets the final slice.

While the space–time approach shows clear advantages up to moderate dimension for European contracts, extending it to American options is an open question as the early–exercise condition induces a complementarity structure that must be enforced at all time layers in the system. For this reason, in the Numerical Experiments Sec. \ref{sec:numerics} we first benchmark our time–stepping solver on European options and use those results as a baseline before moving to American options, where the early–exercise condition can be enforced efficiently after each implicit time step.

\subsection{Greeks on the QTT Grid}

The \emph{Greeks} quantify sensitivities of the option value to market inputs and are central to hedging, risk, and model validation. 
Two commonly used Greeks are Delta, $\Delta=\partial V/\partial S$, which measures the rate of change of the option value with respect to the underlying price, and Gamma, $\Gamma=\partial \Delta/\partial S=\partial^{2}V/\partial S^{2}$, which measures how the Delta itself changes with the underlying price.

After running the space–time or time–stepping QTT solver, we obtain the full option pricing solution grid in QTT form. Computing Greeks can then be done directly on this grid with negligible overhead: apply a low-rank QTT derivative operator and sample the resulting tensors at the desired points, which reduces to fast interpolation on the grid. Appendix~\ref{app:space_time_greeks} provides the full explanation and implementation details.

Using the one-asset setting of the previous subsection ($K=65,\, r=0.08,\, \sigma=0.3,\, T=0.25$ and domain $[K/3,3K]$), we compute Delta and Gamma over the entire grid for both solvers:
\begin{enumerate}
\item build the option price solution grid in QTT (time–stepping or space–time),
\item apply the first- and second-derivative QTT operators,
\item compare against the analytical Black--Scholes Greeks over the full grid.
\end{enumerate}

\begin{table}[h]
  \centering
  \begin{tabular*}{0.47\textwidth}{@{\extracolsep{\fill}}cc|ccc@{}}
    \toprule
    \multicolumn{2}{c}{\textbf{Delta}} & \multicolumn{3}{c}{\textbf{Gamma}} \\
    \cmidrule(l){1-2}\cmidrule(l){3-5}
    \textbf{\shortstack{Grid\\ MAE}} &
    \textbf{\shortstack{Max\\ \phantom{a} $|$Error$|$ \phantom{a}}} &
    \textbf{\shortstack{Grid\\ MAE}} &
    \textbf{\shortstack{Max\\ $|$Error$|$}} &
    \textbf{\shortstack{Run\\ Time(s)}} \\
    \hline
    \midrule
1.1e-03    &    5.3e-03      &    3.3e-04    &      1.2e-03  & 0.002\\
4.0e-04    &     2.1e-03     &     1.0e-04     &     4.5e-04  & 0.003\\
1.5e-04     &    8.8e-04      &    3.4e-05     &     1.9e-04  & 0.005\\
6.2e-05      &   4.7e-04      &    1.9e-05     &     1.5e-04  & 0.007\\
    \hline
    \bottomrule
  \end{tabular*}
  \caption{Grid-wide Delta and Gamma errors and runtime. Rows correspond to Total Cores $=\{10,12,14,16\}$. 
MAE denotes the mean absolute error over the full grid, Max$|$Error$|$ is the absolute maximum error on the grid. 
Run Time(s) is the time to build and apply the MPO once the option price grid is already built.}
  \label{tab:delta_gamma_errors}
\end{table}
As the total number of cores increases, both Delta and Gamma errors decrease steadily (the slightly better accuracy of Gamma is due to a cancellation in the chain rule). Gamma runtimes remain in the millisecond range, negligible compared to building the solution grid. This confirms that, once the QTT option price surface is available, dense Greeks can be produced essentially on demand at minimal additional cost.

\section{Numerical Experiments}\label{sec:numerics}
\subsection{European Basket Put Baseline}\label{sec:numerics_euro}
We begin with the time–stepping QTT solver on European options to establish accuracy/rank baselines and tuning choices that will serve as our starting point before turning to American options. A detailed description of the QTT algorithm used in this test is given in Appendix~\ref{app:d-asset_alg}, here we focus on the practical performance.  

To benchmark our solver we consider a $d$-asset European basket put option whose payoff at maturity is 
\begin{align*}
    \max\left(K_i-\sum_{i=1}^{d} S_T^{(i)},0\right).
\end{align*}
The difficulty in pricing this option is that under Black--Scholes dynamics each underlying follows a geometric Brownian motion, and therefore, are log-normally distributed. As the sum of log-normally distributed random variables is not log-normal, it is not possible to derive an (exact) closed-form representation of the basket price. 
For this reason, we compute the price using a deterministic Gauss--Hermite quadrature over the $d$-dimensional standard normal \cite{luo2014fast}. The resulting price can be expressed as a Gaussian integral that we evaluate with a high-order Gauss–Hermite quadrature to obtain a highly accurate reference value for the error analysis of the QTT solver.
Such solutions are seldom available in practice, since key multi-asset payoffs and market dynamics generally do not permit closed-form formulations or this class of approximations. This emphasizes the need for efficient numerical schemes.

We highlight the flexibility of our solver. Once the Black--Scholes operator was discretized and framed in QTT form, pricing a new contract requires only swapping in the appropriate payoff tensor and, if necessary, adjusting the boundary values, leaving the rest of the algorithm unchanged. Consequently, barrier, max/min, and many other exotic multi-asset options can be handled by adjusting the TT ranks required to approximate these contracts payoff and boundary values and the discretization interval.

\begin{figure}[h]
    \includegraphics[width=\linewidth]{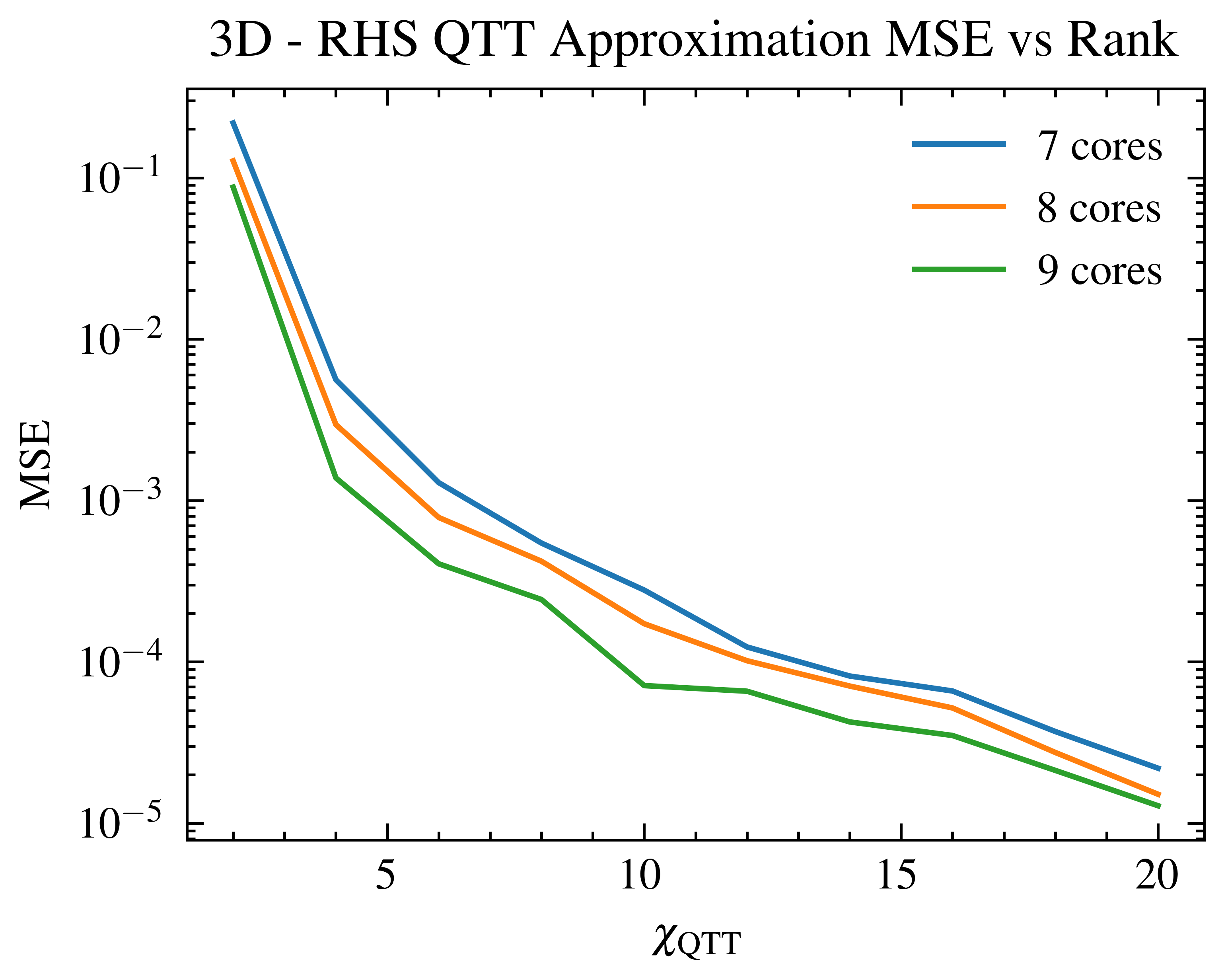}\\[-0.1pt]
    \caption{Rank convergence of the right-hand side approximation for the $d=3$ basket put. MSE versus capped TT-rank $\chi$. Increasing the number of cores per dimension $c$ improves accuracy at fixed $\chi$.}
    \label{fig:rhs_rank_3d_main}
\end{figure}

The solver depends on just two TT rank parameters.
The discretized operator $A$ admits an exact analytic QTT representation with maximal bond dimension $1+(d^2 + 5d)/2$ (see Appendix \ref{app:d-asset_alg}). 
The right-hand side $b$ is built by merging the analytic QTT form of the exponential with a TT-Cross evaluation of the $\max$ function. After the final truncation, capping the TT ranks at $\leq 10$ suffices for the $3$-asset case to achieve the reported accuracy. \blue{Figure}~\ref{fig:rhs_rank_3d_main} 
analyzes the rank convergence of this right-hand side approximation for the $3$-asset case. Appendix~\ref{app:rhs_rank} provides the analogous plot for $d=4$ and discusses these results in more detail. We observe that increasing the number of cores per dimension yields better accuracy at the same bond dimension. Empirically, our experiments suggest that achieving $\mathrm{MSE} \approx 10^{-3}$ for the right-hand side is sufficient to obtain the $2\%$–$1\%$ pricing errors reported in this Section.

We must also choose the spatial domain $[x_{\min},x_{\max}]^{d}$ on which the PDE is solved. Common heuristics set these bounds to a few standard deviations of the underlying log-price or use a payoff-driven criterion to cap the truncation error. In the present study we adopt an adaptive two–stage procedure. First, we run the QTT solver on a coarse grid (only four or five cores per dimension, which is almost cost-free) to obtain a rough approximation of the solution. From this pilot run, we identify the region where the option value is non-negligible and set $x_{\min}$ and $x_{\max}$ accordingly. The final computation is then carried out by doing $2^c$ time steps on a finer grid achieving the desired accuracy without wasting resolution outside the region of interest.

The QTT solver was implemented in Python 3.11 and executed on a MacBook Pro equipped with an Apple M3 Pro processor and 18 GB of RAM. Synthetic numerical values are chosen for market parameters (asset values, interest rates, volatilities and correlations), and for instrument features (strike price, maturity). The numerical values are summarized in Sec. \ref{app:constants}. Table \ref{tab:european} summarizes the wall-clock time required to reach the absolute pricing error against the quadrature reference. 
\begin{table}[h]
\centering
\begin{tabular*}{0.45\textwidth}{@{\extracolsep{\fill}}cccc@{}}
\toprule
\textbf{Assets} &
\textbf{\shortstack{\#Cores per dim}} &
\textbf{Run time(s)} &
\textbf{$\lvert$Error$\rvert$} \\
\hline
        \midrule
        3 & 7 & 16 & $\leq 2\%$\\  
        3 & 8 & 42 & $\leq 1\%$\\  
        4 & 7 & 55 & $\leq 2\%$ \\  
        4 & 8 & 112 & $\leq 1\%$ \\  
        5 & 8 & 198 & $\leq 2\%$ \\  
        5 & 9 & 433 & $\leq 1\%$ \\  
        \bottomrule
    \hline
\end{tabular*}
\caption{Run time and absolute pricing error of our QTT solver for a $d$-asset European basket put.}
\label{tab:european}
\end{table}
It is well known that classical grid–based solvers become impractical once the spatial dimension exceeds three. For instance, repeating the $4$-asset, $8$-core experiment on the same computer would require a grid of $2^{32}$ points; even with an efficient Krylov iteration this would amount to more than a day of run time. The situation deteriorates rapidly: the $5$-asset, $9$-core entry in Table~\ref{tab:european} would involve $2^{45}$ unknowns, so a single solution vector in single precision (float32) would occupy about $128$~TB. No current single-node HPC system offers such RAM, whereas the QTT solver handles the same problem comfortably on a laptop. Although Table~\ref{tab:european} reports the error for a single option price, a key advantage of computing the full solution grid is that one can interpolate prices for any current underlying within the domain without rerunning the solver.

To showcase this capability, we revisit the $d=3$ test with a different strike, $K=33$. With $10$ cores per dimension the spatial grid has $2^{30}$ points; our solver computes the full solution in about $240$ seconds and then prices—via interpolation—options whose baseline spots $S=(10,11,12)$ are all shifted by $\pm10\%$, covering in-the-money/at-the-money/over-the-money cases, with about $2\%$ accuracy across all of them.

\subsection{European Max--Min Put}\label{sec:eur_max_min}
To support the claim that our framework extends to more exotic payoffs with minimal modifications, we repeat the experiment for an European max--min put with payoff
\begin{align*}
    \max(K_i - \min(S_T^{1},\hdots,S_T^{d}),0).
\end{align*}
using the same market parameters as in Table~\ref{tab:european} but with a fixed strike $K=10$. Appendix~\ref{ap:max_min} details the boundary
conditions appropriate for this contract and gives a more in-depth explanation of why, from a PDE viewpoint, the max--min payoff is actually simpler than the basket payoff, allowing us to solve it slightly faster in practice. Table~\ref{tab:eur_max_min} reports the resulting runtimes and pricing errors.

\begin{table}[h]
\centering
\begin{tabular*}{0.45\textwidth}{@{\extracolsep{\fill}}cccc@{}}
\toprule
\textbf{Assets} &
\textbf{\shortstack{\#Cores per dim}} &
\textbf{Run time(s)} &
\textbf{$\lvert$Error$\rvert$} \\
\hline
        \midrule
        3 & 7 & 9 & $\leq 2\%$\\  
        3 & 8 & 23 & $\leq 1\%$\\  
        4 & 7 & 29 & $\leq 2\%$ \\  
        4 & 8 & 61 & $\leq 1\%$ \\  
        5 & 8 & 106 & $\leq 2\%$ \\  
        5 & 9 & 224 & $\leq 1\%$ \\  
        \bottomrule
    \hline
\end{tabular*}
\caption{Run time and absolute pricing error of our QTT solver for a $d$-asset European max--min put.}
\label{tab:eur_max_min}
\end{table}

\subsection{American Basket Put Baseline}\label{sec:numerics_american}

We now turn to the American basket put option. In this type of contract, the holder may exercise at any time $t\in[0,T]$ and receive the payoff: 
$P(S,t) = {\max(K-\sum_{i=1}^{d} S_t^{(i)},0)}$. Because early exercise is allowed, the no-arbitrage condition requires the option value to dominate the payoff at all times, 
$V(S,t) \geq P(S,t)$ for every $t\in[0,T]$, otherwise a risk-free profit would be possible.

Before presenting the pseudocode for this type of contract, we briefly explain the time-stepping update. The full mathematical derivation and QTT constructions are given in Appendices~\ref{app:2-asset} and~\ref{app:d-asset_alg}.
Let $w^i$ denote the QTT approximation of the option value at time level $i$ on the full grid.
Using an implicit discretization, each backward step requires solving a linear system of the form: 
$Aw^{i+1} = b^i$, where $A$ is the (time-independent) FD discretization of the differential elements of the BS PDE and $b^i$ is obtained from the previous iterate after enforcing the boundary conditions.
In Algorithm~\ref{alg:qtt_ts_main}, $\mathrm{A}_{QTT}$ denotes the QTT/MPO representation of $A$,
$w^0=\mathrm{Payoff}_{QTT}$ is the QTT representation of the terminal payoff, and $\mathrm{B}_{QTT}[i]$ is the QTT representation of $b^i$ returned by \texttt{ImposeBoundaries} applied to the current solution. 

\begin{algorithm}[!t]
\caption{QTT time-stepping American option}\label{alg:qtt_ts_main}
\begin{algorithmic}[1]
\Statex{INPUT:} Market Parameters, $d$ = Number of Dimensions,
\Statex{\phantom{INPUT: }}$c$ = Number of Cores p/ dim, $t$ = Time Steps 

\Statex{OUTPUT:} $cd$ cores QTT representation of the solution

\Statex{Build $cd$ cores QTT representation of:}

\State $\textrm{A}_{QTT}$ 

\State $\textrm{Payoff}_{QTT}$ from initial conditions 

\State $w_{sol}[0] \leftarrow \textrm{Payoff}_{QTT}$

\For{$i=0$ until $(t-1)$}

\State $\tau \leftarrow T - i\Delta t$

\State $\textrm{B}_{QTT}[i] \leftarrow$ \texttt{ImposeBoundaries}($w_{sol}[i],\tau$)

\State $w_{sol}[i+1] \leftarrow$ALS($\textrm{A}_{QTT}, w_{sol}[i],\textrm{B}_{QTT}[i]$)

\State $\textrm{Aux}_{QTT} \leftarrow \max(\textrm{Payoff}_{QTT},~w_{sol}[i+1])$

\State $w_{sol}[i+1] \leftarrow \texttt{ReduceRanks}(\textrm{Aux}_{QTT})$

\EndFor
\State Return{} $w_{sol}[t]$
\end{algorithmic}
\end{algorithm}
Algorithm~\ref{alg:qtt_ts_main} summarizes the QTT time-stepping routine used in the experiments below. We note that after each implicit time step, once the linear system has been solved, we enforce the early-exercise condition (EEC) by replacing the tentative price with 
\begin{align}\label{eq:eec}
    V(S,t) \leftarrow \max\!\bigl(V(S,t),\,P(S,t)\bigr).
\end{align}
This element-wise projection can be done with a single TT-Cross (implemented on line 8 of~\ref{alg:qtt_ts_main}). With this same Algorithm the European solver is obtained by omitting lines 8 and 9.

In Table \ref{tab:american} we see that this operation only moderately affects the overall runtime. As shown earlier, TT ranks capped at $12$ already deliver high accuracy. We can efficiently run TT-Cross at a high rank, and truncate them back before the next ALS iteration, keeping both memory usage and computational cost logarithmic in the number of grid points.

For the American basket put option no semi-analytic reference formula is available, so we benchmark the QTT solver against Monte-Carlo regression. Specifically, we implement the Longstaff–Schwartz least-squares algorithm \cite{longstaff2001valuing} with a 95\% confidence interval, that will serve as the reference price against the QTT results.

\begin{table}[h]
\centering
\begin{tabular*}{0.45\textwidth}{@{\extracolsep{\fill}}ccccc@{}}
\toprule
\textbf{Assets} &
\textbf{\shortstack{{\#Cores}\\ {per dim}}} &
\textbf{\shortstack{{Total run}\\ {time(s)}}} &
\textbf{\shortstack{{EEC run}\\ {time(s)}}} &
\textbf{$\lvert$Error$\rvert$} \\
\hline
        \midrule
        3 & 7 & 20 & 5 & $\leq 3\%$\\  
        3 & 8 & 54 & 13 & $\leq 2\%$\\  
        4 & 7 & 71 & 21 & $\leq 3\%$ \\  
        4 & 8 & 143 & 44 & $\leq 2\%$ \\  
        5 & 8 & 253 & 72 & $\leq 3\%$ \\  
        5 & 9 & 558 & 159 & $\leq 2\%$ \\  
        \bottomrule
    \hline
\end{tabular*}
\caption{Run time of our QTT solver for a $d$-asset American basket put with the same market parameters as the previous experiment. EEC run time is the extra run time necessary to impose the early-exercise condition via TT-Cross.}
\label{tab:american}
\end{table}

Overall the early-exercise check adds only a modest overhead. Relative to the European timings in Table \ref{tab:european} the total run time rises by roughly 30\%. For the four and five-asset cases we increased the TT-Cross sweeps used to enforce the early-exercise condition—from three to four to preserve the target accuracy, yet this stage still accounts for no more than one-third of the total runtime. Hence, even for more intricate pay-offs or tighter error tolerances, the extra cost is still expected to scale controllably.

\subsection{American Max--Min Put}
We now consider the American counterpart of the max--min contract, using the same market parameters as in the European experiments and a fixed strike $K = 10$. In contrast to the European case, as shown in Table~\ref{tab:ame_max_min}, this option is slower to price. Although the boundary conditions are actually simpler for the max--min payoff, the evaluation of the early-exercise condition on the tentative solution and the payoff becomes more expensive: we require a higher bond dimension to faithfully represent the QTT form of the payoff, which directly increases the cost of the TT-Cross projection step. A detailed analysis of the rank behavior for this contract is given in Appendix~\ref{ap:max_min}.
\begin{table}[h]
\centering
\begin{tabular*}{0.45\textwidth}{@{\extracolsep{\fill}}cccc@{}}
\toprule
\textbf{Assets} &
\textbf{\shortstack{\#Cores per dim}} &
\textbf{Run time(s)} &
\textbf{$\lvert$Error$\rvert$} \\
\hline
        \midrule
        3 & 7 & 31 & $\leq 2\%$\\  
        3 & 8 & 83 & $\leq 1\%$\\  
        4 & 7 & 115 & $\leq 2\%$ \\  
        4 & 8 & 233 & $\leq 1\%$ \\  
        5 & 8 & 430 & $\leq 2\%$ \\  
        5 & 9 & 953 & $\leq 1\%$ \\  
        \bottomrule
    \hline
\end{tabular*}
\caption{Run time and absolute pricing error of our QTT solver for a $d$-asset American max--min put.}
\label{tab:ame_max_min}
\end{table}

\section{Discussion}
This work shows that quantized tensor trains (QTT) enable deterministic full-grid solutions of the multi-asset Black–Scholes PDE beyond the classical three-dimensional limit on a PC. By representing the operator, payoffs, and boundary conditions in QTT form with ranks that scale polynomially in the number of assets and remain independent of the grid size, we obtain solvers whose memory and runtime growth are fundamentally different compared to dense or sparse grid methods. In contrast to Monte Carlo approaches, the proposed framework recovers the entire solution grid and smooth Greeks without statistical noise.

A key empirical observation is the favorable behavior of the QTT ranks. For basket and max–min options, in industrially relevant regimes, the maximal ranks increase moderately from three to five dimensions and remain stable under grid refinement. Even for American options, where early-exercise constraints introduce nonlinearity, the rank growth remains controlled, making large full-grid computations feasible.

The comparison between time-stepping and space–time formulations highlights complementary trade-offs. Time-stepping achieves lower ranks, while the space–time approach computes the entire solution in a single solve. Both methods naturally provide Greeks, virtually for free, through differentiation of the tensor representation. Which formulation is preferable depends on the number of assets and the rank growth of the payoff and boundary conditions.

In contrast to classical grid-based solvers, the bottleneck in QTT is not the number of discretization points but the tensor ranks required to represent the payoff and boundary conditions. This shifts the problem from “how fine can the grid be?” to “how efficiently can we encode the non-smooth nonlinearities that drive rank growth?” Our current pipeline, combining analytic constructions with TT-Cross, was sufficient for all experiments reported here, but it is not the only option. A promising next step is to compare alternative interpolation/cross-approximation strategies that may further reduce ranks and evaluation cost, such as tensor cross  \cite{nunez2025learning} and interpolative \cite{lindsey2023multiscale} constructions for QTT representations of payoffs and boundary conditions. 

Beyond the constant-coefficient Black–Scholes setting, the same framework can be extended to richer market models. Natural targets include local-volatility models \cite{dupire1994pricing}, where the instantaneous volatility becomes a deterministic function of spot and time, and jump–diffusion extensions such as Merton’s model \cite{merton1976option} or SABR \cite{hagan2014arbitrage}. For these more complex models, the QTT framework could directly benefit from pairing its low-rank representation with more sophisticated finite-difference discretizations (e.g., operator splitting/ADI, see \cite{duffy2006fdm} for a practical overview). 

Overall, these results position QTT as a particularly compelling deterministic route to full-grid prices and Greeks in genuinely high-dimensional Black–Scholes settings: when the entire solution surface is required, classical PDE solvers become grid-limited and Monte Carlo remains intrinsically sample-limited, whereas QTT is the only approach we found that keeps full-grid computation tractable beyond three dimensions on a PC.

\paragraph{Acknowledgments}
We thank Martin Mikkelsen for a careful reading of the manuscript.
We acknowledge support from the Carlsberg foundation and the Novo Nordisk Foundation. 

~ \newpage
\bibliography{main}
\bibliographystyle{apsrev4-1}

~ \newpage ~ \newpage 
\onecolumngrid

\appendix

\section{Algorithms}\label{appen:Algorithms}

This appendix presents the algorithms from the main paper in ascending order of complexity, so the reader can progress from the simplest setting to the most general with minimal conceptual overhead. For the $d$-asset space–time and time-stepping solvers, we also provide their time-complexity analyses. In addition, we show how to compute the Greeks directly from the QTT grid produced by our solvers. Although we frame the discussion in the classical Black--Scholes model, the same constructions extend to richer market models and, more broadly, to other parabolic PDEs beyond option pricing.

\subsection{Space--Time QTT Solver for the 1-Asset Black--Scholes PDE}\label{appen:space_time_1D}

In this section, we build the Algorithm presented in Section \ref{sec:space_time_main} of the main paper to solve the single-asset Black--Scholes PDE using a space--time approach. Our goal is to reformulate the PDE in a way to obtain a low-rank QTT representation of the differential elements.

Consider the Black--Scholes PDE for $0 < t < T$ and $S>0$:

\begin{align}\label{eq:Original_1DBS}
    \frac{\partial V}{\partial t} + 
    \frac{1}{2}\sigma^2 S^2 \frac{\partial^2 V}{\partial S^2} + 
    r S \frac{\partial V}{\partial S}  
     - rV = 0,
\end{align}
where $V = V(t, S)$ denotes the price of the option at time $t$ with asset price $S$. The volatility of $S$ is given by $\sigma$ and $r$ is the risk-free interest rate.
To eliminate the explicit dependence on $S$, let $x = \ln S$. Applying the chain rule transforms \eqref{eq:Original_1DBS} into:
\begin{align}\label{eq:log1DBS}
    \frac{\partial V}{\partial t} + 
    \frac{\sigma^2}{2} \frac{\partial^2 V}{\partial x^2} + 
    \left(r - \frac{\sigma^2}{2}\right) \frac{\partial V}{\partial x} - rV = 0.
\end{align}
Next, we reverse time using the substitution $\tau = T - t$, so that $\tau = 0$ corresponds to maturity and $\tau = T$ is the current time. Under this change of variable, the previous PDE  becomes:
\begin{align}\label{eq:1DBS}
    \frac{\partial V}{\partial \tau} - 
    \frac{\sigma^2}{2} \frac{\partial^2 V}{\partial x^2} - 
    \left(r - \frac{\sigma^2}{2}\right) \frac{\partial V}{\partial x} + rV = 0.
\end{align}
For a European call option with strike $K$, the terminal condition (initial condition in $\tau$) is given by:
\begin{align}
    V(\tau = 0, x) = \max\left(e^x - K, 0\right).
\end{align}
At the spatial boundaries, we impose:
\begin{align*}
    V(\tau, x_{\min}) &= 0  &\text{(option is worthless)}, \\
    V(\tau, x_{\max}) &= e^{x_{\max}} - K e^{-r \tau}  &\text{(deep in-the-money)}.
\end{align*}
These conditions are enforced on a truncated domain $(\tau,x)  \in [0, T] \times [x_{\min}, x_{\max}]$.

We now discretize the transformed PDE \eqref{eq:1DBS} on a uniform grid. Let $\tau_i = i\,\Delta t$, for $i = 0, \dots, N$ with $\Delta t = T/N$, and $x_j = x_{\min} + j\,\Delta x$, for $j = 0, \dots, M+1$ with $\Delta x = (x_{\max} - x_{\min})/(M+1)$. Let $w^i_j \approx V(\tau_i,x_j)$ denote the approximation of the solution on the discrete grid ordered serially when flattened into a vector (see Eq. \eqref{eq:spacetime_serial}) and consider the following discretization of \eqref{eq:1DBS} for the interior nodes $i=1,\dots,N$ and $j=1,\dots,M$:
\begin{align}\label{eq:discrete_BS}
    \frac{w^i_j - w^{i-1}_j}{\Delta t} 
    - \frac{\sigma^2}{2} \frac{w^i_{j+1} - 2w^i_j + w^i_{j-1}}{\Delta x^2}
    - \left(r - \frac{\sigma^2}{2}\right) \frac{w^i_{j+1} - w^i_{j-1}}{2\Delta x}
    + r w^i_j = 0. 
\end{align}
For simplicity, let $\alpha = \sigma^2/(2\Delta x^2)$ and $\beta = \left(r - \sigma^2/2\right)/(2\Delta x)$. We can write Equation \eqref{eq:discrete_BS} as:
\begin{align}\label{eq:C0_C+_C-}
    \underbrace{(1+2\alpha \Delta t + r \Delta t)}_{C_0} w^i_j ~ \underbrace{- \Delta t (\alpha+\beta)}_{C_+} w^i_{j+1} ~ 
    \underbrace{- \Delta t (\alpha-\beta)}_{C_{-}} w^i_{j-1} - w^{i-1}_j = 0.
\end{align}
Each time layer $i$ couples only through the spatial stencil defined by coefficients $C_0$, $C_+$, and $C_-$, and connects to the previous layer via $-\I$, where $\I$ is the identity matrix. Thus, the fully discrete system can be expressed as a block lower-bidiagonal system $Aw = b$:
\begin{align*}
A =\begin{pmatrix}
\mathrm{B} &  &        \\
-\I & \mathrm{B} &        \\
& -\I & \mathrm{B} &        \\
& &\ddots & \ddots
\end{pmatrix}_{NM \times NM},
B=
\begin{pmatrix}
C_0 & C_+ &        \\
C_- & C_0 & C_+        \\
& C_- & C_0 & C_+        \\
& &\ddots & \ddots & \ddots
\end{pmatrix}_{M \times M},
\end{align*}
where $\I$ is the $M \times M$ identity matrix, and the solution vector is given by
\begin{align}\label{eq:spacetime_serial}
    w = [w^1_1,w^1_2,\dots,w^1_{M},w^2_1,\dots,w^2_{M},\dots \dots,w^N_1,\dots,w^N_{M} ]^\T.
\end{align}

To efficiently solve the system $Aw = b$, we now move to the QTT framework. First we construct a rank-4 QTT representation of the matrix $A$. Let $N = 2^n$ and $M = 2^m$ for any integers $n, m \ge 2$, and define the matrix $L_{N \times N}$ with $-1$ in the lower subdiagonal and zeros elsewhere.  
Then $A$ can be written as a Kronecker sum:
\begin{align*}
    A = L_{N \times N} \otimes \I_{M \times M} + 
    \I_{N \times N} \otimes B_{M \times M},
\end{align*}
by Lemma~\ref{lemma_1}, both $L$ and $B$ have an exact low-rank QTT representation with bond dimension 3 thus for any $n$ and $m$ the matrix $A$ have an exact QTT representation of rank 4.

It remains to construct the QTT representation of the vector $b$, which encodes the terminal condition (initial in $\tau$) and boundary conditions at $x = x_{\min}$ and $x = x_{\max}$.  
Using the tensor product structure of the space–time grid, we express $b$ as the following sum:
\begin{align}\label{eq:b_vec}
    b &= | 0 \rangle^n \otimes w^0_j -
        C_{-} w^i_0 \otimes | 0 \rangle^m - 
        C_{+} w^i_{M+1} \otimes | 1 \rangle^m
      = | 0 \rangle^n \otimes \underbrace{\max(e^{x_j} - K,0)}_{\text{payoff}} -
        C_{+} \underbrace{(e^{x_{\text{max}}}-K e^{-r \tau})}_{\text{deep ITM}} \otimes | 1 \rangle^m,
\end{align}
where $| 0 \rangle^n$ and $| 1 \rangle^m$ are the vectors of length $2^n$ and $2^m$ of the form $(1,0,\hdots,0)^\T$ and $(0,0,\hdots,1)^\T$, respectively that have an exact rank-1 QTT representation and $w^i_0=0$ for the call option. As mentioned in the main paper, the payoff term can be built using different methods e.g., TT-Cross or TT-SVD and the deep in-the-money term admits an exact rank-2 analytic QTT representation (see Appendix \ref{appen:analytic_qtt_exp}), as it is the product of exponentials in separable variables. Hence, we can construct a precise and low-rank QTT representation of $b$.

After constructing the QTT representations of $A$ and $b$, we use them as input for the (M)ALS solver to obtain an approximate solution $w$. To obtain the option price at a given underlying value $S$, set $x=\ln S$ and interpolate the final-time slice $w^N_{1:M}$ on the spatial grid $\{x_j\}_{j=1}^{M}$ at $x$.

Although we focus here on option pricing, the space–time formulation is advantageous whenever the entire solution surface of a parabolic PDE is required not merely its terminal values making the approach broadly relevant to other problems in financial mathematics or other fields.

\subsection{Computing the Greeks from the QTT Grid}\label{app:space_time_greeks}

Let $\mathrm{QTT}_w$ be the output of the space–time algorithm for the 1-asset Black--Scholes PDE (App.~\ref{appen:space_time_1D}). 
It is a $2c$ cores (wlog assuming $c_x=c_t=c$ cores per dimension) QTT representation of the vector
\begin{align}\label{eq:v_space-time}
    v = [v^1_1, v^1_2, \dots, v^1_{2^c},\; v^2_1, \dots, v^2_{2^c},\; \dots,\; v^{2^c}_1, \dots, v^{2^c}_{2^c}]^\T,
\end{align}
our discrete approximation to $V(\tau,x)$, the solution of the transformed BS PDE \eqref{eq:log1DBS}, where $\tau=T-t$ and $x=\ln S$. 
The last $2^c$ entries form the spatial solution at the final time; this slice coincides with running $2^c$ implicit steps of size $1/2^c$ in the time–stepping scheme. 
We compute spot Greeks from this terminal slice. 
For notational convenience, set
$$
  v^{2^c}_{1:2^c}=(v^{2^c}_1,\dots,v^{2^c}_{2^c})^\T \approx V(x_1,\dots,x_{2^c}),
$$
where $\{x_j\}$ are the spatial grid points.

By the chain rule, the standard Black--Scholes Greeks relate to derivatives in $x$ under the transformed PDE as
\begin{align*}
    \Delta &= \frac{\partial V}{\partial S}
           = \frac{1}{S}\,\frac{\partial V}{\partial x},\\
    \Gamma &= \frac{\partial^2 V}{\partial S^2}
           = \frac{1}{S^2}\!\left(\frac{\partial^2 V}{\partial x^2}
                                  - \frac{\partial V}{\partial x}\right).
\end{align*}

\paragraph{QTT operators.}
To obtain $\partial V/\partial x$ from $\mathrm{QTT}_w$ we apply the QTT operator
$$
  \mathcal{D}_x \;=\; I_{c_t} \otimes D_x,
$$
where $I_{c_t}$ is the identity on the $c_t$ time cores and $D_x$ is the first–derivative MPO on the $c_x$ space cores (exact and low-rank). 
In QTT terms this is implemented by concatenating the $c_t$ identity cores with the $c_x$ derivative cores and applying the resulting $2c$ cores MPO to $\mathrm{QTT}_w$. 
For the time–stepping output, we simply apply $D_x$ to the $c_x$ spatial cores, no identity concatenation is needed.

\paragraph{Delta.}
Let $\widehat{V}_x = \mathcal{D}_x \mathrm{QTT}_w$ denote the QTT approximation to $\partial V/\partial x$ at the terminal time. To evaluate $\Delta(S_i)$, that is, the delta of an option whose asset value is $S_i=e^{x_i}$ we interpolate $\widehat{V}_x$ at $x_i$ on the QTT grid and rescale:
$$\Delta(S_i) \;\approx\; \frac{1}{S_i}\,\widehat{V}_x(x_i).$$

\paragraph{Gamma.}
First, build the MPO representation of the second derivative $D_{xx}$ (also exact and low-rank). Similarly to Delta, let ${\widehat{V}_{xx} \;=\; (I_{c_t}\!\otimes D_{xx})\,\mathrm{QTT}_w}$. Then interpolate at $x_i$ and combine:
$$
  \Gamma(S_i) \;\approx\; \frac{1}{S_i^{2}}\Bigl(\widehat{V}_{xx}(x_i)-\widehat{V}_x(x_i)\Bigr).
$$

\paragraph{Remarks.}
(i) All derivative MPOs above admit analytic QTT constructions with small bond dimension (see Lemma~\ref{lemma_1}) on dyadic grids, so the cost is a single MPO–MPS application. 
(ii) The multi-asset extension inserts identities on all non–differentiated spatial cores, e.g.\ $\Delta_i:\; D_{x_i}\bigotimes_{j\neq i} I_{x_j}$, preserving low ranks under the Kronecker structure.

\subsection{Time Complexity of the Space--Time QTT Solver for the d-Asset Black--Scholes PDE}\label{appen:space_time_analysis}

In this section, we derive an upper bound on the time complexity of the multi-asset space--time QTT solver used in Section \ref{sec:space_time_main}. We start with two assets, verify the TT-rank growth, and then state the general $d$-asset result.

As presented in Section \ref{sec:multi_asset_option} after the log–transform ($x=\ln S_1$, $y=\ln S_2$) the two-asset Black--Scholes PDE is
\begin{align}
\partial_t V = \frac{\sigma^2_x}{2} \partial_{xx}V +
                \frac{\sigma^2_y}{2} \partial_{yy}V +
                \rho \sigma_x \sigma_y\partial_{xy}V+
                (r - \frac{\sigma^2_x}{2}) \partial_x V +
                (r - \frac{\sigma^2_y}{2}) \partial_y V -
                rV.
\end{align}
Before moving to the $d$-asset case we build the equivalent $A$ matrix for this two-asset case. By doing the same steps presented in Eqs: \eqref{eq:log1DBS} to \eqref{eq:C0_C+_C-} with the added mixed derivative term a fully implicit space--time stencil yields, for interior nodes,
\begin{align*}
C_1 w_{j,k}^{i}
+C_2 w_{j+1,k}^{i}+C_3 w_{j-1,k}^{i}
+C_4 w_{j,k+1}^{i}+C_5 w_{j,k-1}^{i}
-C_{xy}\!\bigl[w_{j+1,k+1}^{i}-w_{j+1,k-1}^{i}-w_{j-1,k+1}^{i}+w_{j-1,k-1}^{i}\bigr]
-w_{j,k}^{\,i-1}=0,
\end{align*}
where the coefficients $C_1,\dots,C_5,C_{xy}$ are defined in a similar way to Equation \eqref{eq:C0_C+_C-}. By the same argument of the one-asset case, the fully discrete system can be expressed as the system of linear equations $Aw=b$. To build $A$ we need first to define the following matrices:
\begin{align*}
B &=B_x+B_y-B_{xy},\\
B_x &=\text{tridiag}(C_3,C_1,C_2)\otimes \I,\\
B_y &=\I\otimes\text{tridiag}(C_5,0,C_4),\\
B_{xy}&=C_{xy}\bigl(U\otimes(U-L)+L\otimes(L-U)\bigr),
\end{align*}
where $U_{ij}=\delta_{i+1,j}$ and $L_{ij}=\delta_{i,j+1}$.
The matrix $A$ have the same block-bidiagonal structure as in the one-asset case but now has size $NM_{x}M_{y} \times NM_{x}M_{y}$. By Lemma~\ref{lemma_1} $B$ has a QTT representation of rank: $\chi_{B_{x}} + \chi_{\I}+\chi_{B_{xy}} =3+1+6=10$, and hence $\chi_A\le 10+1=11$. The last $1$ comes from the QTT representation of $\I$ on the diagonal of $A$.
For $3$-assets we have 
$B=B_x+B_y+B_z-(B_{xy}+B_{xz}+B_{yz})$,
so $\chi_B\le 3+1+1+6+6+1=18$. By induction, we can generalize this argument for $d$-assets:
\begin{align*}
    \chi_A \leq \frac{d^2 + 13d-8}{2}.
\end{align*}

To extend the right–hand side $b$ to $d$ spatial dimensions we first note that, even if we are interested in pricing a call option, it is often cheaper to compute the corresponding put and then convert it through the put–call parity. Under the boundary convention adopted in Appendix~\ref{app:d-asset_alg}, that every upper face is fixed to zero, half of the $2d$ boundaries vanish, so the put formulation is more economical in QTT form.

For the general $b$ vector the only source of rank growth is the nonlinear $\max$ term, which we approximate with TT-Cross. All remaining factors are exponentials and admit analytic rank-1 QTT representations. Numerical experiments show that, even after the TT-Cross step, the ranks of $b$ stay modest while high accuracy is preserved. Hence, the cost of building the QTT representation of $b$ scales polynomially in the number of cores $c$ and the dimension $d$, rather than exponentially in the full grid size $2^{cd}$.

Combining the rank estimates for $A$ and $b$ with the per sweep cost of ALS, the $d$-asset space–time solver delivers the full grid of $2^{cd}$ values at each of the $2^{c}$ time levels in
\begin{align}
    \mathcal{O}\!\bigl(c(d+1)\, [\tfrac{1}{2}(d^{2}+13d-8)]^{2}\, \chi_{b}^{3}\, 4\,\gamma\bigr) = \mathcal{O}\!\bigl( (cd^{5}+cd^3)\,\chi_{b}^{3}\bigr),
\end{align}
where $4$ is the square of the mode size and $\gamma$ is the cost of the local solver. The two-core MALS variant raises only the mode size to the power of three instead of two.

\subsection{Time-Stepping QTT Solver for the 2-Asset Black--Scholes PDE}\label{app:2-asset}

In this section, we develop a standard time-stepping finite difference scheme for the two-asset Black--Scholes PDE and reformulate it in the QTT framework. Our goal is to provide a clear mapping from the classical finite difference formulation to its QTT counterpart. This construction will be extended to the $d$-asset case in Section \ref{app:d-asset_alg} and could also serve as a starting point to solve similar PDEs arising in financial mathematics and related fields.

Generalizing the one-asset Black--Scholes equation \eqref{eq:log1DBS} to two underlying assets $S_1$ and $S_2$, we obtain the following log-transformed PDE for the option price $V = V(x, y, t)$, where $x = \ln S_1$, $y = \ln S_2$, and $t \in [0, T]$:
\begin{align}\label{eq:Original_2DBS}
\partial_t V = \frac{\sigma^2_x}{2} \partial_{xx}V +
                \frac{\sigma^2_y}{2} \partial_{yy}V +
                \rho \sigma_x \sigma_y\partial_{xy}V+
                (r - \frac{\sigma^2_x}{2}) \partial_x V +
                (r - \frac{\sigma^2_y}{2}) \partial_y V -
                rV,
\end{align}
Here, $\rho \in [-1, 1]$ is the correlation between the two assets. The other terms are similar to the one-asset case and we don't reverse the time direction keeping $0 < t < T$. 

For a European Basket call option with strike $K$, the terminal condition is given by:
\begin{align}\label{eq:payoff_basket2d}
    V(x,y,T) = \max (\omega_1 e^x + \omega_2 e^y - K, 0),
\end{align}
w.l.o.g we assume unit weights $\omega_1$ and $\omega_2$. The spatial boundary conditions are chosen to match asymptotic behavior:
\begin{align}\label{eq:bcs_basket}
  \begin{aligned}
   &V(x_{\min},y,t) = \max( e^y - K e^{-rt} , 0),\\       
   &V(x_{\max},y,t) = e^{x_{\max}} + e^y - K e^{-rt},
  \end{aligned}
  &&
  \begin{aligned}
   &V(x,y_{\min},t) = \max( e^x - K e^{-rt} , 0), \\       
   &V(x,y_{\max},t) = e^x + e^{y_{\max}} - K e^{-rt}.
  \end{aligned}
 \end{align}
We discretize Equation \eqref{eq:Original_2DBS} in the same uniform grid as the one presented in Section \ref{appen:space_time_1D} and define $y_k$ to have the same structure of $x_j$ with $\Delta y=(y_{\max} - y_{\min})/(M+1)$. Let $w^i_{j,k} \approx V(x_j, y_k, t_i)$ be an approximation of the solution such that at every time step the 2D grid is flattened into a vector (using serial ordering see Eq. \eqref{eq:spacetime_serial}) and consider a general $\theta$-scheme in time, with $\theta \in [0,1]$, yielding:
\begin{align*}
    \frac{w^i_{j,k} - w^{i+1}_{j,k}}{\Delta t} 
    = \theta \, \mathcal{L} w^i_{j,k} + (1 - \theta)\, \mathcal{L} w^{i+1}_{j,k},
\end{align*}
where $\mathcal{L}$ is the spatial differential operator from the right-hand side of \eqref{eq:Original_2DBS}. Rearranging the terms gives:
\begin{align}
    \left(\I - \Delta t\theta \mathcal{L} \right) w^i_{j,k}
    = \left(\I + \Delta t(1 - \theta) \mathcal{L} \right) w^{i+1}_{j,k}, \label{eq:theta_scheme}
\end{align}
where $w^i_{j,k}$ is the solution at time level $t_i$, and $\I$ denotes the identity matrix.
This general time-stepping scheme reduces to the fully implicit method when $\theta = 1$, to the Crank–Nicolson method when $\theta = 0.5$, and to the explicit Euler method when $\theta = 0$. Among these, the fully implicit scheme simplifies this equation to a standard linear system of the form:
\begin{align}
    A w^i_{j,k} = w^{i+1}_{j,k}, \label{eq:implicit_system}
\end{align}
where $A = \I - \Delta t \mathcal{L}$ encodes the discretized spatial operator. Since our goal in the QTT framework is usually to construct low-rank representations of both operators and solution vectors, the fully implicit scheme will be our method of choice.

Now we show that the operator $\mathcal{L}$ can be written as a Kronecker sum. Let $D_x$ and $D_{xx}$ denote the first and second-order centered finite-difference matrices on a uniform grid with spacing $\Delta x$.
Both are Toeplitz tridiagonal matrix that can be expressed in stencil form as $D_x = \frac{1}{2 \Delta x}[-1,0,1]$ and $D_{xx} = \frac{1}{\Delta x^2}[1,-2,1]$. 
Because the same grid is used in the $y$-direction, we set $D_{y}=D_{x}$ and $D_{yy}=D_{xx}$. Then by a straightforward calculation we have the following:
\begin{align}
\mathcal{L} &=
      \frac{\sigma_x^{2}}{2}\,(D_{xx}\otimes \I)
    + \frac{\sigma_y^{2}}{2}\,(\I\otimes D_{yy})
    + \rho\,\sigma_x\sigma_y\,(D_x\otimes D_y)
    + \Bigl(r-\frac{\sigma_x^{2}}{2}\Bigr)\,(D_x\otimes \I)
    + \Bigl(r-\frac{\sigma_y^{2}}{2}\Bigr)\,(\I\otimes D_y)
    - r\,(\I\otimes \I) \notag \\[4pt]
&=  \underbrace{\frac{\sigma_x^{2}}{2}\,(D_{xx}\otimes \I)
   + \Bigl(r-\frac{\sigma_x^{2}}{2}\Bigr)\,(D_x\otimes \I)
   - r\,(\I\otimes \I)}_{(1,3,\hdots,3,1;1,1,\hdots,1,1)}
   + \underbrace{\frac{\sigma_y^{2}}{2}\,(\I\otimes D_{yy})
   + \Bigl(r-\frac{\sigma_y^{2}}{2}\Bigr)\,(\I\otimes D_y)}_{(1,1,\hdots,1,1;1,3,\hdots,3,1)}
   - \underbrace{\rho\,\sigma_x\sigma_y\,(D_x\otimes D_y)}_{(1,3,\hdots,3,1;1,3,\hdots,3,1)}
   \label{eq:L_operator}.
\end{align} 
If each of the three one-dimensional matrices has size $2^{c}\times 2^{c}$, Lemma~\ref{lemma_1} guarantees the QTT ranks indicated by the number sequences beneath the three blocks in \eqref{eq:L_operator}. Hence the maximal rank of $\mathcal{L}$ is bounded by the sum
\begin{align}\label{eq:ineq}
\operatorname{rank}_{\text{QTT}}(\mathcal{L})
\;\le\; 3 + 1 + 3\;=\; 7,
\end{align}
so $\mathcal{L}$ admits an \emph{exact} QTT representation of rank at most 7, independent of the grid resolution. Because the time-stepping matrix is $A=\I-\Delta t\mathcal{L}$, adding a rank-1 identity term raises the bound by one, giving $\operatorname{rank}_{\text{QTT}}(A)\le 8$.

The main steps of the time-stepping algorithm to solve the European two-asset Black--Scholes equation are given below:

\begin{algorithm}[H]
\caption{QTT Time-Stepping Solver for the 2D European Black--Scholes Equation}\label{alg:2D_TS_BS}
\begin{algorithmic}[1]
\Statex{INPUT:} $c$, \textit{timesteps}, T, K, r, $\sigma_x$, $\sigma_y$, $\rho$

\Statex{OUTPUT:} $2c$ cores QTT representation of the solution

\Statex{Build $2c$ cores QTT representation of (1,2):}

\State $\textrm{A}_{QTT}$ \Comment{Eq. \eqref{eq:L_operator} and construction from Lemma~\ref{lemma_1}}

\State $\textrm{Payoff}_{QTT}$ from initial conditions \Comment{Eq. \eqref{eq:payoff_basket2d} using construction from \ref{appen:analytic_qtt_exp} and TT-Cross}

\State $w_{sol}[0] \leftarrow \textrm{Payoff}_{QTT}$

\For{$i=0$ until (\textit{timesteps}-1)}

\State $\tau \leftarrow T - i\Delta t$

\State $\textrm{B}_{QTT}[i] \leftarrow$ \texttt{impose\underline{~}boundaries}($w_{sol}[i],\tau$)

\State $w_{sol}[i+1] \leftarrow$(M)ALS($\textrm{A}_{QTT}, w_{sol}[i],\textrm{B}_{QTT}[i]$)

\EndFor
\State Return{} $w_{sol}[\textit{timesteps}]$
\end{algorithmic}
\end{algorithm}
\noindent This algorithm is highly efficient since all steps in the main loop are performed entirely within the QTT framework. The primary computational cost comes from running ALS over the required number of time steps. In the next Section we will derive an upper bound for the time complexity of this algorithm. In Sec.~\ref{sec:numerics_american} we present a small modification to this algorithm to price American options.

The \texttt{impose\underline{~}boundaries} function receives the current solution and overwrites the four boundary values according to the conditions in Equations \eqref{eq:bcs_basket}. This update can be carried out entirely in the QTT framework by applying a MPO to the QTT representation of the solution. The first step of this function is to zero all the elements in the four boundaries. We start by defining the following vectors:
\begin{align}\label{eq:v_l_and_v_lr}
    \mathbf{1}_{2^c} = 
    \begin{bNiceMatrix}
    1 \\ 1 \\ \Vdots \\ 1 \\ 1
\end{bNiceMatrix}_{2^c}\textrm{~,~}
v_{\text{L}} = 
    \begin{bNiceMatrix}
    0 \\ 1 \\ \Vdots \\ 1 \\ 1
\end{bNiceMatrix}_{2^c} \textrm{~and~}
v_{\text{R}} =\begin{bNiceMatrix}
    1 \\ 1 \\ \Vdots \\ 1 \\ 0
\end{bNiceMatrix}_{2^c}.
\end{align}
Let $v_{\textrm{eraser}} = 
(v_{\text{L}} \otimes \mathbf{1})
(\mathbf{1} \otimes v_{\text{L}})
(v_{\text{R}} \otimes \mathbf{1})
(\mathbf{1} \otimes v_{\text{R}})
$. Multiplying this vector to the solution vector sets all boundary values to zero while leaving all the interior entries unchanged.
To build the QTT representation of these vectors we use construction \ref{appen:v_eraser}.

The second step inserts the four boundary values into the system of linear equations we are solving. Using bra-ket notation and setting $\ketzero$ and $\ketone$ as the vectors of length $c$ of the form $(1,0,\hdots,0)$ and $(0,0,\hdots,1)$, respectively we have the following expression to add these values to match the indexing of the solution vector:
\begin{align*}
    \textrm{Boundary}_{QTT} = (\ketzero \otimes \textrm{L}_{QTT}) + 
        (\textrm{D}_{QTT} \otimes \ketzero)  +
        (\ketone \otimes \textrm{R}_{QTT}) +
        (\textrm{U}_{QTT} \otimes \ketone),
\end{align*}
where $\textrm{L,D,R,U}_{QTT}$ are the $c$ cores QTT representation of the left, down, right and up boundary conditions given in~\eqref{eq:bcs_basket} that are computed with the appropriate value of $t$. To build the QTT representation of the left and down boundary conditions, we use TT-Cross and for the other two boundary conditions we use the analytic QTT representation of the exponential function (App. \ref{appen:analytic_qtt_exp}).

Finally, putting all these two steps together the action of the \texttt{impose\underline{~}boundaries} on the QTT representation of the solution vector $w_{sol}$ is:
\begin{align*}
    \textrm{B}_{QTT} = 
    (v_{\textrm{eraser}})_{QTT} . w_{sol} + \textrm{Boundary}_{QTT},
\end{align*}
where the first term zeros the four boundary faces and the second term adds the correct boundary values.

\subsection{Time-Stepping QTT Solver for the d-Asset Black--Scholes PDE}\label{app:d-asset_alg}

This section extends the QTT-based finite-difference algorithm from the $2$-asset case (App. \ref{app:2-asset}) to the general $d$-asset case. We highlight only the elements that differ considerably from the 2-asset formulation.

Generalizing the log-transformed Black--Scholes equation to $d$ correlated underlyings $S_1,\dots,S_d$, we set $x_i=\ln S_i$ and denote the option price by 
$V = V(x_1,\dots,x_d,t)$ for $t\in[0,T]$. As in the 2-asset case, upon discretization at every time-step the $d$-dimensional grid is flattened into a single vector using serial ordering. Let $\sigma_i>0$ be the volatility of $S_i$ and $\rho_{ij}\in[-1,1]$ the instantaneous correlation between $S_i$ and $S_j$ $(i\neq j)$.
The resulting $d$-dimensional Black--Scholes PDE is  
\begin{align*}
\partial_t V \;=\;
\frac{1}{2}\sum_{i=1}^{d}\sigma_i^{2}\,\partial_{x_i x_i}V
\;+\;
\frac{1}{2}\sum_{\substack{i,j=1\\i\neq j}}^{d}
      \rho_{ij}\,\sigma_i\sigma_j\,\partial_{x_i x_j}V
\;+\;
\sum_{i=1}^{d}\Bigl(r-\tfrac{\sigma_i^{2}}{2}\Bigr)\partial_{x_i}V
\;-\;
r\,V.
\end{align*}
The spatial operator $\mathcal{L}$ extends to $d$ underlyings exactly as in the two-asset case.  
For three assets, Eq.~\eqref{eq:L_operator} contains six Kronecker blocks. By Lemma~\ref{lemma_1},
\begin{align*}
\operatorname{rank}_{\text{QTT}}(\mathcal{L})
   \;\le\; 3 + 1 + 1 + 3 + 3 + 1 \;=\; 12.
\end{align*}
By induction, for four assets the same counting gives $\operatorname{rank}_{\text{QTT}}(\mathcal{L}) \;\le\; 18$. For $d$-assets
$\operatorname{rank}_{\text{QTT}}(\mathcal{L}) \;\le\; d\,(d+5)/2$,
so $\mathcal{L}$ admits an exact QTT representation of low rank even in high dimensions.

For a European $d$-asset Basket put option with strike $K$, the terminal condition is given by:
\begin{align*}
    V(x_1,\hdots,x_d,T) = \max (K - (\omega_1 e^{x_1} + \cdots +  \omega_d e^{x_d}) , 0),
\end{align*}
w.l.o.g we assume unit weights $w_i$. The spatial boundary conditions are chosen to match asymptotic behavior:
\begin{align}
V(x_{1\min},x_2,\hdots,x_d,t) &= \max(Ke^{-rt} - (e^{x_2} + e^{x_3}+\cdots + e^{x_d}),0),\label{eq:d-asset_bc_1}\\
V(x_{1},x_{2\min},\hdots,x_d,t) &= \max(Ke^{-rt} - (e^{x_1} + e^{x_3}+\cdots + e^{x_d}),0),\label{eq:d-asset_bc_2}\\
&\phantom{..}\vdots \nonumber \\
V(x_{1},x_{2},\hdots,x_{d\min},t) &= \max(Ke^{-rt} - (e^{x_1} + e^{x_2}+ \cdots + e^{x_{d-1}}),0)\label{eq:d-asset_bc_d}.
\end{align}
We impose homogeneous Dirichlet conditions on every upper boundary. Whenever $x_i = x_{i\max}$ for any $1 \leq i \leq d$, we set
\begin{align}\label{eq:upper_BC}
   V(x_1,\dots,x_{i\max},\dots,x_d,t)=0.
\end{align}

Now we extend the \texttt{impose\underline{~}boundaries} function  to an arbitrary number of spatial dimensions $d$. First we set to zero all the elements on any of the $2 \cdot d$ boundary hyper-planes. This is accomplished by applying the \texttt{Eraser\underline{~}MPO} and then imposing the $d$ values given by Equations \eqref{eq:d-asset_bc_1} to \eqref{eq:d-asset_bc_d} to the QTT representation of the system of linear equations we are solving.

To build this \texttt{Eraser\underline{~}MPO} first we build the $c$ cores QTT representation of $\mathbf{1}_{2^c}, v_{\text{L}}$ and $v_{\text{R}}$ as defined in \eqref{eq:v_l_and_v_lr}, where $c$ is the number of cores per dimension. Then we build the following $2 \cdot d$ auxiliary MPOs:
\begin{align*}
    MPO_{x_1\min} &= v_{\text{L}} \otimes \underbrace{\mathbf{1}_{2^c} \otimes \mathbf{1}_{2^c} \otimes \cdots \otimes \mathbf{1}_{2^c}}_{d-1}\\
    MPO_{x_2\min} &= \mathbf{1}_{2^c} \otimes \underbrace{v_{\text{L}} \otimes \mathbf{1}_{2^c} \otimes \cdots \otimes \mathbf{1}_{2^c}}_{d-1}\\
    &\phantom{..}\vdots\\
    MPO_{x_d\min} &= \mathbf{1}_{2^c} \otimes \underbrace{\mathbf{1}_{2^c} \otimes \cdots \otimes \mathbf{1}_{2^c} \otimes v_{\text{L}}}_{d-1},
\end{align*}
and $$MPO_{x_i\max} = \underbrace{\mathbf{1}_{2^c} \otimes \cdots \otimes \mathbf{1}_{2^c} \otimes \overbrace{v_{\text{R}}}^{i~\text{position}}\otimes \mathbf{1}_{2^c} \otimes \cdots \otimes \mathbf{1}_{2^c}}_{d~\text{terms}}, ~\text{for~} 1 \leq i \leq d.$$
Finally, by multiplying all of the $2 \cdot d$ terms together we have the following definition:
\begin{align*}
    \text{\texttt{Eraser\underline{~}MPO}} = \prod_{i=1}^d MPO_{x_i\min} \cdot MPO_{x_i\max}.
\end{align*}

Now the second step of the \texttt{impose\underline{~}boundaries} function is to insert the boundary values into the system of linear equations we are going to solve. For ease of notation we detail the four-dimensional case. Noting that this construction extends naturally to any number of dimensions.

Let $c$ be the number of cores for each one of the four dimensions. To impose the boundary condition given by Equation \eqref{eq:d-asset_bc_1} we first build a $3c$ cores analytical QTT representation of the function $({Ke^{-rt} - (e^{x_2} + e^{x_3}+e^{x_4})})$ using the construction give in Appendix~$\ref{appen:analytic_qtt_exp}$. We then apply TT-Cross to replace negative entries by zero ($\max$ function). Finally, we append a $c$ cores QTT representation of $\ketzero$ to the \emph{leftmost} end of the resulting $3c$ cores QTT.

To impose the boundary condition in Eq.~\eqref{eq:d-asset_bc_d} for the four-dimensional case, we follow the same steps with the corresponding boundary expression, but append $\ketzero$ to the \emph{rightmost} end of the $3c$ cores QTT.

For the other two boundary conditions instead of concatenating a $\ketzero$ to the beginning or end of a $3c$ cores QTT we concatenate $c$ \textit{square} cores (drawn as black squares in the figure) to a position corresponding to the $x_{i \min}$ variable. As an example, to impose the boundary condition in Eq.~\eqref{eq:d-asset_bc_2} for the four-dimensional case we first build a $3c$ cores QTT representation of the function $(Ke^{-rt} - (e^{x_1} + e^{x_3}+ e^{x_4}))$ as shown in the left-hand QTT in the diagram below. Next we insert $c$ \textit{square} cores at the position corresponding to the $x_{2 \min}$ register (right–hand QTT). In braket notation the \textit{square} cores inserts $c$ zero bits in the second register.\\

\hspace{1.1cm}%
\tikz[baseline=-0.5ex]{
    \node[draw, circle, inner sep=0.5pt]                        (A1)  {\phantom{\scriptsize q}};
    \node[right=0.31cm of A1]                                   (Adots1) {$\cdots$};
    \node[draw, circle, inner sep=0.5pt, right=0.31cm of Adots1](A4)  {\phantom{\scriptsize q}};
    \node[draw, circle, inner sep=0.5pt, right=0.21cm of A4]    (A5)  {\phantom{\scriptsize q}};
    \node[right=0.31cm of A5]                                   (Adots2) {$\cdots$};
    \node[draw, circle, inner sep=0.5pt, right=0.31cm of Adots2](A9)  {\phantom{\scriptsize q}};
    \node[draw, circle, inner sep=0.5pt, right=0.21cm of A9]    (A10) {\phantom{\scriptsize q}};
    \node[right=0.31cm of A10]                                  (Adots3) {$\cdots$};
    \node[draw, circle, inner sep=0.5pt, right=0.31cm of Adots3](A12) {\phantom{\scriptsize q}};

    \node[right=0.8cm of Adots3] (imp) {$\implies$};

    \draw (A1) -- (Adots1) -- (A4);
    \draw (A4) -- (A5);
    \draw (A5) -- (Adots2) -- (A9);
    \draw (A9) -- (A10);
    \draw (A10) -- (Adots3) -- (A12);

    \node[rectangle, draw=none, minimum size=1pt] at (1.85, 0.2) {\scriptsize$\chi$};

    \foreach \n/\col in {A1/red,A4/red,A5/nicegreen,A9/nicegreen,A10/blue,A12/blue}{
        \draw (\n) -- ++(0,-0.4);
        \node[below] at ([yshift=-0.3cm]\n) {\scriptsize\textcolor{\col}{2}};
    }
}%
%
\hspace{0.1cm}%
%
\tikz[baseline=-0.5ex]{
    \node[draw, circle, inner sep=0.5pt]                        (B1)  {\phantom{\scriptsize q}};
    \node[right=0.31cm of B1]                                   (Bdots1) {$\cdots$};
    \node[draw, circle, inner sep=0.5pt, right=0.31cm of Bdots1](B4)  {\phantom{\scriptsize q}};
    \node[draw, rectangle, inner sep=0.5pt, right=0.21cm of B4]    (B4a) {\phantom{\scriptsize q}};
    \node[right=0.21cm of B4a]                                  (Bmid) {$\cdots$};
    \node[draw, rectangle, inner sep=0.5pt, right=0.21cm of Bmid]  (B4b) {\phantom{\scriptsize q}};
    \node[draw, circle, inner sep=0.5pt, right=0.21cm of B4b]   (B5)  {\phantom{\scriptsize q}};
    \node[right=0.31cm of B5]                                   (Bdots2) {$\cdots$};
    \node[draw, circle, inner sep=0.5pt, right=0.31cm of Bdots2](B9)  {\phantom{\scriptsize q}};
    \node[draw, circle, inner sep=0.5pt, right=0.21cm of B9]    (B10) {\phantom{\scriptsize q}};
    \node[right=0.31cm of B10]                                  (Bdots3) {$\cdots$};
    \node[draw, circle, inner sep=0.5pt, right=0.31cm of Bdots3](B12) {\phantom{\scriptsize q}};

    \draw (B1)  -- (Bdots1) -- (B4);
    \draw (B4)  -- (B4a) -- (Bmid) -- (B4b) -- (B5);
    \draw (B5)  -- (Bdots2) -- (B9);
    \draw (B9)  -- (B10);
    \draw (B10) -- (Bdots3) -- (B12);

    \node[rectangle, draw=none, minimum size=1pt] at (1.85, 0.2) {\scriptsize$\chi$};
    \node[rectangle, draw=none, minimum size=1pt] at (2.3, 0.2) {\scriptsize$\chi$};
    \node[rectangle, draw=none, minimum size=1pt] at (3.2, 0.2) {\scriptsize$\chi$};
    \node[rectangle, draw=none, minimum size=1pt] at (3.65, 0.2) {\scriptsize$\chi$};

    \foreach \n/\col in {B1/red,B4/red,B4a/black,B4b/black,B5/nicegreen,B9/nicegreen,B10/blue,B12/blue}{
        \draw (\n) -- ++(0,-0.4);
        \node[below] at ([yshift=-0.3cm]\n) {\scriptsize\textcolor{\col}{2}};
    }
}

\begin{align*}
    \hspace{-0.5cm}\sum_{b \in \{0,1\}^{3c}} \alpha_b 
    |\underbrace{b_{3c} \cdots b_{2c+1}}_{\textcolor{red}{c}}  
    \underbrace{b_{2c} \cdots b_{c+1}}_{\textcolor{nicegreen}{c}} 
    \underbrace{b_c \cdots b_1}_{{\textcolor{blue}{c}}} \rangle 
    \hspace{0.26cm}\implies 
    \sum_{b \in \{0,1\}^{3c}} \alpha_b 
    |b_{3c} \cdots b_{2c+1} 
    \underbrace{0 0 \cdots 0}_{c} 
    b_{2c} \cdots b_{c+1} 
    b_c \cdots b_1 \rangle.
\end{align*}
The square cores 
\begin{tikzpicture}[baseline=-0.5ex]
  \node[draw, rectangle, inner sep=3.5pt] (S) {};
  \draw (S.west) -- ++(-0.4cm,0) node[midway,above=0.8pt] {\scriptsize $\chi$};

  \draw (S.east) -- ++(0.4cm,0) node[midway,above=0.8pt] {\scriptsize $\chi$};

  \draw (S.south) -- ++(0,-0.3cm);
  \node[below] at ([yshift=-0.35cm]S) {\scriptsize $2$};
\end{tikzpicture} $= \mathrm{S}^{\chi,2,1,\chi}$ are defined as: 
$\mathrm{S}^{i,j,k,l} = $
$\begin{cases}
  1, & \text{if } j=k=0 \text{ and } 0 \leq i=l \leq r-1, \\[4pt]
  0,  & \text{otherwise }.
\end{cases}$

\noindent The final step to impose this boundary condition is to apply TT-Cross in the $4c$ cores QTT (right-hand side of the diagram) to replace the negative entries by zero ($\max$ function).

For the remaining boundary, ${V(x_{1},x_{2},x_{3 \min},x_4,t) = \max(Ke^{-rt} - (e^{x_1} + e^{x_2}+e^{x_4}),0)}$ the construction is identical except that the block of square cores is inserted at the $x_{3\min}$ register, yielding the following QTT:
\tikz[baseline=-0.5ex]{
    \node[draw, circle, inner sep=0.5pt]                        (B1)  {\phantom{\scriptsize q}};
    \node[right=0.31cm of B1]                                   (Bdots1) {$\cdots$};
    \node[draw, circle, inner sep=0.5pt, right=0.31cm of Bdots1](B4)  {\phantom{\scriptsize q}};
    \node[draw, circle, inner sep=0.5pt, right=0.21cm of B4]    (B4a) {\phantom{\scriptsize q}};
    \node[right=0.21cm of B4a]                                  (Bmid) {$\cdots$};
    \node[draw, circle, inner sep=0.5pt, right=0.21cm of Bmid]  (B4b) {\phantom{\scriptsize q}};
    \node[draw, rectangle, inner sep=0.5pt, right=0.21cm of B4b]   (B5)  {\phantom{\scriptsize q}};
    \node[right=0.31cm of B5]                                   (Bdots2) {$\cdots$};
    \node[draw, rectangle, inner sep=0.5pt, right=0.31cm of Bdots2](B9)  {\phantom{\scriptsize q}};
    \node[draw, circle, inner sep=0.5pt, right=0.21cm of B9]    (B10) {\phantom{\scriptsize q}};
    \node[right=0.31cm of B10]                                  (Bdots3) {$\cdots$};
    \node[draw, circle, inner sep=0.5pt, right=0.31cm of Bdots3](B12) {\phantom{\scriptsize q}};

    \draw (B1)  -- (Bdots1) -- (B4);
    \draw (B4)  -- (B4a) -- (Bmid) -- (B4b) -- (B5);
    \draw (B5)  -- (Bdots2) -- (B9);
    \draw (B9)  -- (B10);
    \draw (B10) -- (Bdots3) -- (B12);


    \foreach \n/\col in {B1/red,B4/red,B4a/nicegreen,B4b/nicegreen,B5/black,B9/black,B10/blue,B12/blue}{
        \draw (\n) -- ++(0,-0.4);
        \node[below] at ([yshift=-0.3cm]\n) {\scriptsize\textcolor{\col}{2}};
    }
}. The final step of the \texttt{impose\underline{~}boundaries} function is to sum these four QTTs and use them to solve the next iteration of the system of linear equations as we did on the two-asset case (see line 6 of Algorithm \ref{alg:2D_TS_BS}), before advancing to the next time step.

To extend \texttt{impose\_boundaries} to $d$ spatial dimensions with $c$ cores per dimension we note that the construction still falls into two categories independently of $d$:

\begin{enumerate}
\item Equations \eqref{eq:d-asset_bc_1} and \eqref{eq:d-asset_bc_d}:
      Build a QTT with $c(d-1)$ cores for the unfixed variables and
      append a $c$ cores representation of $\ketzero$ to the
      far left (for $x_{1\min}$) or to the far right (for $x_{d \min}$) of this initial QTT.

\item Remaining $d-2$ equations:
      Start from the $c(d-1)$ cores QTT of the unfixed variables and insert a block of $c$ square cores at the register associated with the fixed coordinate.
\end{enumerate}
We note that before applying the TT-Cross operation, the $4c$-core QTT on the right-hand side of the diagram still has a small TT-rank~$\chi$. Indeed, it is merely a sum of exponentials, each of which admits an analytic rank-1 QTT representation, so the combined bond dimension remains low. After the application of TT-Cross we can efficiently cap the TT-rank and still obtain the desired accuracy without increasing the overall run time (See App. \ref{app:rhs_rank} for a study on the RHS rank convergence).

Putting all these elements together, the overall time complexity of the $d$-asset time-stepping QTT solver, where each asset has $c$ cores and we perform $2^c$ time steps, is $\mathcal{O}\bigl((\text{ALS}+\texttt{impose\_boundaries})\cdot \text{TimeSteps}\bigr)$. By the discussion above, the cost of $\texttt{impose\_boundaries}$ is $\mathcal{O}\bigl(d\cdot(\text{Eraser\_MPO}+\mathrm{TT\text{-}Cross})\bigr)$ and we denote it by $\beta$. One sweep of ALS costs $\mathcal{O}\bigl(c d\,\chi_A^{2}\,\chi_b^{3}4\,\gamma\bigr)$, where $4$ is the square of the mode size and $\gamma$ is the cost of the local solver. Using the bounds derived in this section, the overall time complexity is
\begin{align*}
    \mathcal{O}\Bigl(\bigl(c d\,\bigl[\tfrac{1}{2}(d^{2}+5d)\bigr]^{2}\,\chi_b^{3} 4\,\gamma\,\beta\bigr)\,2^{c}\Bigr),
\end{align*}
which, keeping only the dominant terms, simplifies to
$$\mathcal{O}\bigl(2^{c}\,\bigl(c d^{5}+c d^{3}\bigr)\,\chi_b^{3}\bigr).$$
We end this section with the observation that, from a practical standpoint, optimizing two cores at a time (MALS) during the initial time steps improves accuracy at a modest runtime trade-off.

\section{Supplementary Data and Experiments}

\subsection{Basket RHS rank convergence}\label{app:rhs_rank}
This subsection documents the empirical rank behavior of the right-hand side (RHS) used in our time-stepping $d$-asset Black--Scholes QTT solver (App.~\ref{app:d-asset_alg}), with market parameters as in Section~\ref{app:constants}. 
The RHS combines the payoff: $\max(K_i-\sum_{i=1}^{d} S_T^{(i)},0)$ with spatial boundary terms enforced at every time step (Eqs:\eqref{eq:d-asset_bc_1} to \eqref{eq:d-asset_bc_d}).

\paragraph{Experimental setup.}
We fix the number of QTT cores per spatial dimension to $c\in\{7,8,9\}$. 
For each $d\in\{3,4\}$ we:
(a) build the analytic QTT representation for the exponential terms;
(b) apply TT-Cross to evaluate the $\max$ term;  
(c) truncate the result by capping the TT ranks at $\chi_{\mathrm{QTT}}$ (with some target tolerance $\varepsilon$); and
(d) compute the MSE against a dense reference of the same vector obtained by sampling a $2^{15}$ grid points on all tests.

\begin{figure}[h]
  \centering
  \begin{minipage}[t]{0.4\linewidth}
    \centering
    \includegraphics[width=\linewidth]{Figures/3D_chi_new.png}\\[-0.1pt]
  \end{minipage}\hspace{0.06\linewidth}
  \begin{minipage}[t]{0.4\linewidth}
    \centering
    \includegraphics[width=\linewidth]{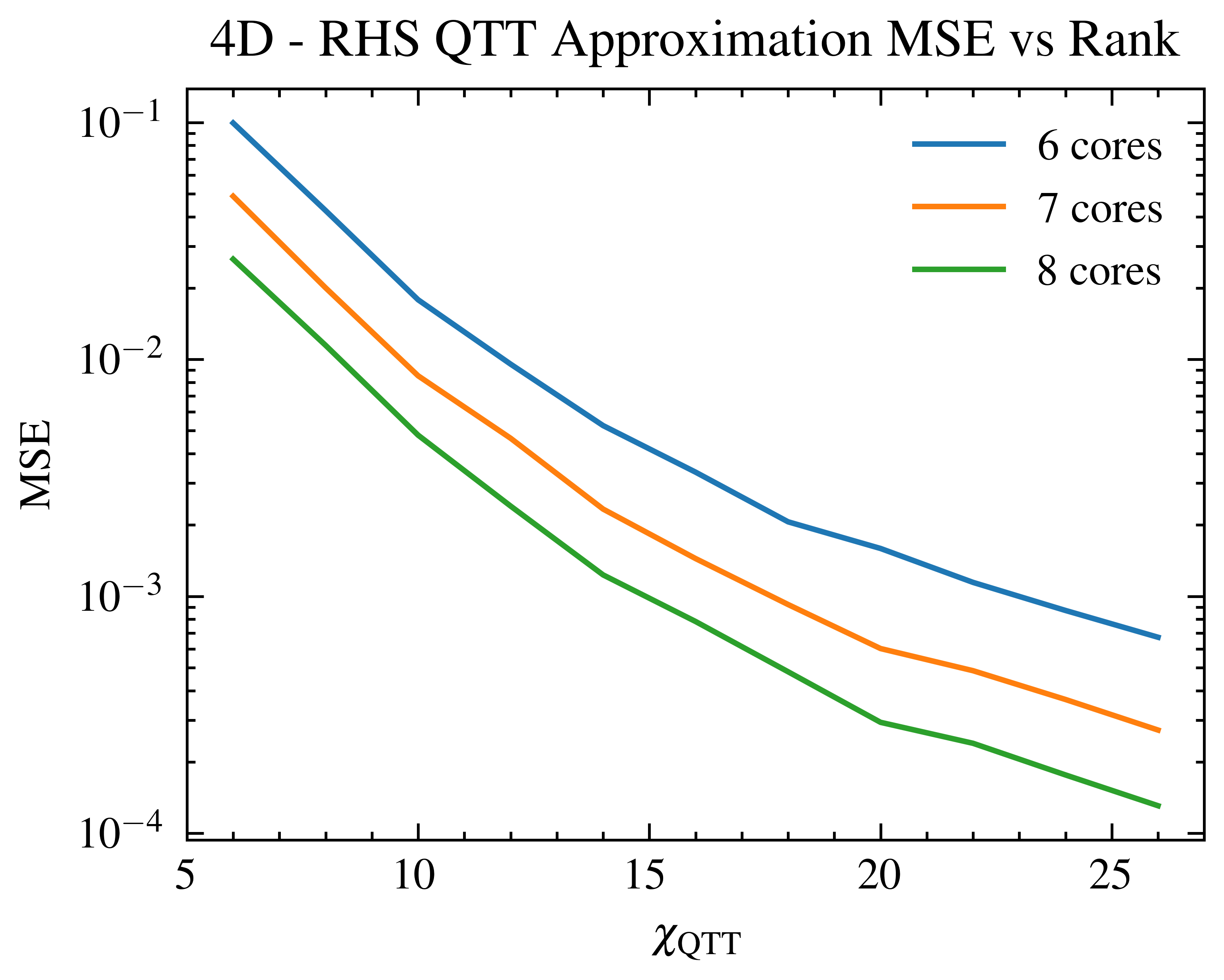}\\[-0.1pt]
  \end{minipage}
  \label{fig:rhs_side_by_side}
\end{figure}

\paragraph{Takeaway.}
As detailed in App.~\ref{app:d-asset_alg}, constructing the spatial boundaries in $d$ dimensions requires $d$ TT-Cross evaluations. Empirically, we note that an RHS accuracy of $\mathrm{MSE} \approx 10^{-3}$ is sufficient to obtain the $2\%$–$1\%$ pricing errors reported in the experiments of the main text.

\subsection{Choice of constants for the experiments in Sec. \ref{sec:numerics}}\label{app:constants}

We specify the synthetic market inputs once as ordered vectors and then reuse the appropriate prefixes for each one of the $3,4$ or $5$-asset test case:\\
spot prices:
$S=(10,11,12,13,14)$,\\
strikes:
$K=(34,47,62)$,\\
volatilities and correlations
\begin{equation}
\sigma = \left(\begin{matrix}
    0.25\\0.15\\0.20\\0.10\\0.15
\end{matrix}\right),~~~~~~~~~~~
    \rho= \left(\begin{matrix}
        1 &0.4 &0.3 &0.2 &0.1\\
        0.4 & 1 & 0.2 & 0.3 & 0.4\\
        0.3&0.2&1&0.1&0.2\\
        0.2&0.3&0.1&1&0.3\\
        0.1&0.4&0.2&0.3&1
    \end{matrix}\right),
\end{equation}
maturity $T=1.0$ and risk-free interest rate $r=5\%$.
For a $d$-asset experiment we simply take the first $d$ entries of each list.
For example, the 3-asset case uses
${(S_1,S_2,S_3)=(10,11,12)}$, strike ${K=34}$,
${(\sigma_1,\sigma_2,\sigma_3)=(0.25,0.15,0.20)}$,
and correlations ${(\rho_{12},\rho_{13},\rho_{23})=(0.4,0.3,0.2)}$,
with the common $T$ and $r$. The 4- and 5-asset tests are formed analogously.

\subsection{3-asset max--min put (worst-of put)}\label{ap:max_min}

We adapt the time–stepping QTT solver for the basket option (Ap. \ref{app:d-asset_alg}) to the $3$-asset European put on the minimum and examine the main difference between these two contracts.
For a European $3$-asset (put on the minimum) with strike $K$, the terminal condition (payoff) is given by 
\begin{align}\label{eq:payoff_maxmin}
V(x_1,x_2,x_3,T)
  &= \max\!\Bigl(K - \min\!\bigl(e^{x_1},e^{x_2},e^{x_3}\bigr),\,0\Bigr).
\end{align}
The spatial boundary conditions reflect the fact that, as soon as one asset becomes very small, it is effectively the minimum and the option behaves like a discounted bond. We impose on the lower boundaries:
\begin{align}\label{eq:lower_bc_maxmin}
V(x_{1\min},x_2,x_3,t) = V(x_1,x_{2\min},x_3,t) = V(x_1,x_2,x_{3\min},t) =K e^{-rt}.
\end{align}
For the upper boundaries we reuse the homogeneous Dirichlet condition from the basket case (Eq.~\ref{eq:upper_BC}). Alternatively, we could also enforce homogeneous Neumann conditions 
\begin{equation*}
\frac{\partial V}{\partial x_i}(x_1,\dots,x_{i,\max},\dots,x_d,t)=0\quad (1\le i\le d),
\end{equation*}
for a higher number of assets $(d \ge 5)$ this will be more appropriate. 

As mentioned in Section \ref{sec:eur_max_min} for European contracts this type of option is actually simpler than the basket case. We do not need to apply the non-linear $\max$ at every time step to enforce the boundary conditions (Eqs. \eqref{eq:lower_bc_maxmin}), and the spatial boundary terms admit an explicit rank-1 QTT construction. 
The only change is in the terminal payoff, which introduces an additional $\min$; for Europeans this payoff is built once at $t=T$, so even if its intermediate TT ranks are higher, the one-off cost does not impact overall runtime. 
Empirically, this is confirmed by our benchmarks (see Table \ref{tab:eur_max_min}): the $d$-asset worst-of put is solved slightly faster than the $d$-asset basket put under the same market parameters as Section~\ref{sec:numerics_euro}, using $K=10$ to yield a realistic near-ATM configuration.

For American contracts, the early-exercise condition (Eq.~\eqref{eq:eec}) must be enforced at every time step. In practice, this remains efficient provided the QTT payoff in Eq.~\eqref{eq:payoff_maxmin} is kept low-rank and achieves RHS accuracy $\mathrm{MSE}<10^{-2}$, which is sufficient for our $1–2\%$ pricing targets. To verify this, we build the 3-asset payoff in QTT with the same market parameters as above and $c=8$ cores per dimension, realizing the $\min$ nonlinearity via TT-Cross and then truncate the rank at the final TT-Cross application . Accuracy is assessed against a dense $256^3$ reference. The figure below fixes $S_3$ at the mid index and displays: QTT payoffs at two rank caps (12 and 24) and the pointwise squared error, with the global MSE reported in the title.

\begin{figure}[h]
  \centering
  \begin{minipage}[t]{0.85\linewidth}
    \centering
    \includegraphics[width=\linewidth]{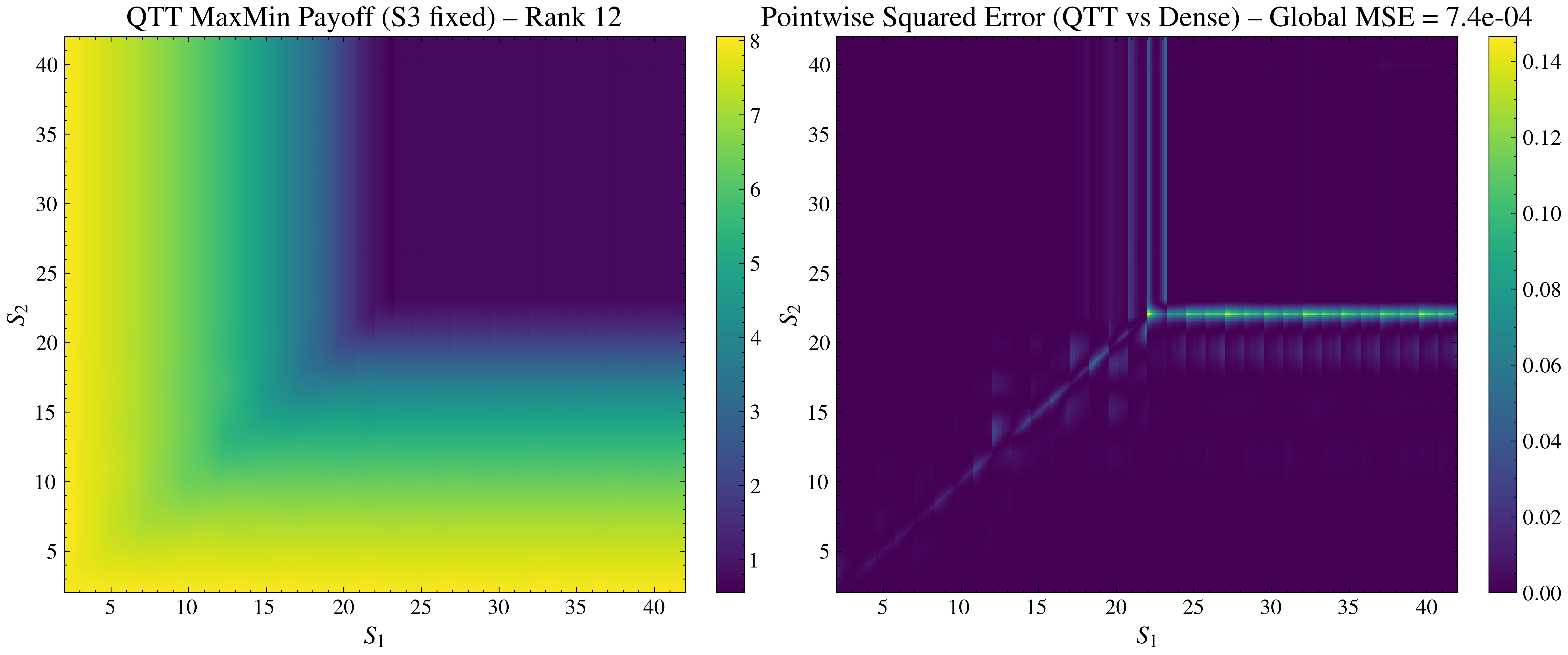}\\[-0.1pt]
  \end{minipage}\hspace{0.001\linewidth}
  \begin{minipage}[t]{0.85\linewidth}
    \centering
    \includegraphics[width=\linewidth]{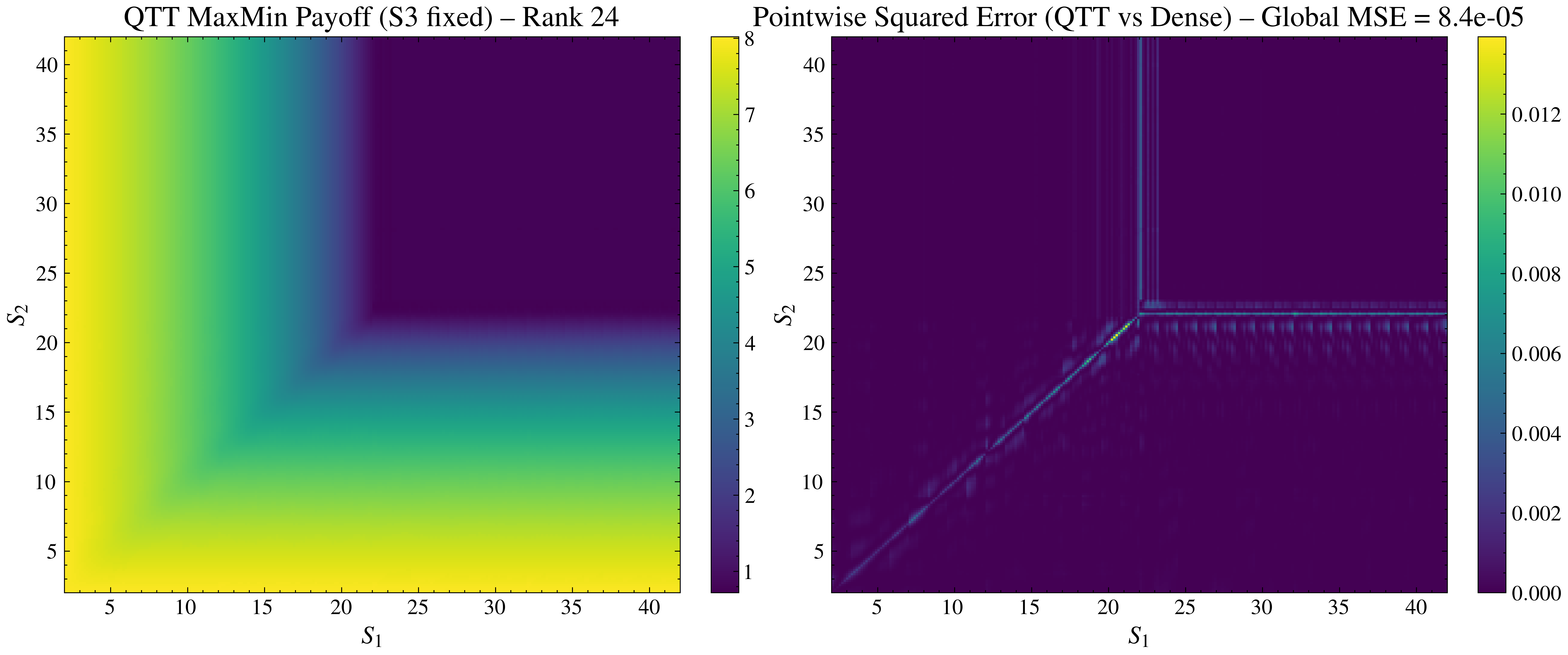}\\[-0.1pt]
  \end{minipage}
  \caption{Dense vs. QTT payoff and pointwise squared error for the 3-asset worst-of put on a
  $(256)^{\times 3}$ grid (slice at fixed $S_3$). The QTT payoff is built via TT-Cross
  for $\min(S_1,S_2,S_3)$ and $\max(\cdot,0)$, then we truncate the rank at $12$ and $24$.
  As expected the error concentrates near the exercise kink but can be reduced by increasing the truncation rank.}
  \label{fig:worst_of_put}
\end{figure}
\noindent The dense and QTT slices coincide away from the kink, with discrepancies localized along $\min(S_1,S_2,S_3)\approx K$. Increasing the rank cap tightens the error band. Unlike the European case, the American worst-of put runs slightly slower than the basket option (see Table~\ref{tab:ame_max_min}), although the runtime remains at the same order of magnitude.

\section{Useful QTT Constructions}
\label{ap:Usefull_QTT_Const}

In this section, we present key constructions for building QTT representation of vectors and matrices that are utilized in the algorithms discussed in this paper. We used the same indexing and notation as the one presented in Section \ref{sec:tt}.

\subsection{Building the Analytic QTT Representation of \texorpdfstring{$f(x)=e^{\alpha x}$}{f(x)=exp(alpha x)}}\label{appen:analytic_qtt_exp}

We construct the analytic rank-1 QTT representation of $f(x) = e^{\alpha x}$, where $x$ is discretized in the interval $(0,1)$ with $2^c$ grid points. Using our indexing convention, the QTT representation of the discretized function $f$ function is given by:
\begin{align*}
  \F_1^{(1,2,1,1)} \bowtie \F_2^{(1,2,1,1)} \bowtie \cdots \bowtie \F_c^{(1,2,1,1)},
\end{align*}
where the tensor components are defined as:
\begin{align*}
    F_i^{0,:,0,0} &= 
    \begin{bmatrix} 
    1 \\ \exp(\alpha x[2^{c-i}])
    \end{bmatrix}.
\end{align*}

\subsubsection{Adjusting to another interval}

Although the above construction assumes that $x$ is discretized in $(0,1)$ with $2^c$ points, our solvers typically require values in another discrete interval, for example, $(S_{min}, S_{max})$, where these values refer to the minimum and maximum values of the stock for a given problem. To construct the analytic QTT representation of $g(y) = e^y$, $y \in (a,b)$ first construct the QTT representation of $f(x) = e^{(b-a)x}$ as shown above and multiply by $e^{a}$ to get the QTT representation of $e^y$.

\subsubsection{Higher Dimensions}

Extending the analytical QTT representation from the 1D case to the 2D function $f(x,y)=e^{\alpha x}\,e^{\beta y}$ is immediate, because the 2D function factorizes as a Kronecker product of its 1D components. In practice, we first build the QTT representation of each exponential factor, using the desired number of cores and the appropriate spatial interval, and then concatenate these tensors to obtain the serial QTT representation of the discretized 2D function.  
The same construction generalizes recursively to any number of spatial dimensions.

\subsection{Building the QTT representation of \texorpdfstring{$v_{\text{L}}$}{v\_l} and \texorpdfstring{$v_{\text{R}}$}{v\_r}}\label{appen:v_eraser}

For any $c \geq 2$ we define the following vectors:
\begin{align*}
    v_{\text{L}} = 
    \begin{bNiceMatrix}
    0 \\ 1 \\ \Vdots \\ 1 \\ 1
\end{bNiceMatrix}_{2^c} \textrm{~and~~~}
v_{\text{R}} = 
    \begin{bNiceMatrix}
    1 \\ 1 \\ \Vdots \\ 1 \\ 0
\end{bNiceMatrix}_{2^c}.
\end{align*}
The following construction builds the rank-$2$ QTT representation of $v_{\text{L}}$ with $c$ cores:
\begin{align*}
v_{\text{L}}=\F^{1,2,1,2}
 \bowtie
  (\M^{2,2,1,2})^{\bowtie (c-2)} \bowtie
  \mathrm{L}^{2,2,1,1},
\end{align*}
with all entries of $\F$ and $\M$ equal to zero except: 
${\F_{0,0,0,0}=1},~  {\F_{0,1,0,1}=1},~{\M_{0,0,0,0}=1,}~{\M_{0,1,0,1}=1},~{\M_{1,0,0,1}=1},$ ${\M_{1,1,0,1}=1}$ and all entries of $\mathrm{L}$ equal to one except for $\mathrm{L}_{0,0,0,0}=0$, and the following construction builds the rank-$2$ QTT representation of $v_{\text{R}}$ with $c$ cores:
\begin{align*}
v_{\text{R}}=\F^{1,2,1,2}
 \bowtie
  (\M^{2,2,1,2})^{\bowtie (c-2)} \bowtie
  \mathrm{L}^{2,2,1,1},
\end{align*}
with all entries of $\F$ and $\M$ equal to zero except: 
${\F_{0,0,0,0}=1},~  {\F_{0,1,0,1}=1},~{\M_{0,0,0,0}=1,}~{\M_{0,1,0,0}=1},~{\M_{1,0,0,0}=1},$ ${\M_{1,1,0,1}=1}$ and all entries of $\mathrm{L}$ equal to one except for $\mathrm{L}_{1,1,0,0}=0$.

\subsection{Building the QTT representation of the Toeplitz tridiagonal matrix}\label{appen:lemma_1}
\begin{lemX}[\cite{kazeev2012}]\label{lemma_1}
Let $I = 
\begin{psmallmatrix}
1 & 0\\
0 & 1
\end{psmallmatrix}$, 
$J = 
\begin{psmallmatrix}
0 & 1\\
0 & 0
\end{psmallmatrix}$, 
$J' = 
\begin{psmallmatrix}
0 & 0\\
1 & 0
\end{psmallmatrix}$, and $\alpha, \beta, \gamma\in \mathbb{C}$, then for any integer 
$c \geq 2$, the $2^c \times 2^c$ matrix 
\begin{align*}
D_{\alpha,\beta,\gamma} = 
\begin{pmatrix}
\alpha & \beta &  &  &    \\
\gamma & \alpha & \beta &  &    \\
 & \ddots & \ddots & \ddots &    \\
\end{pmatrix}
\end{align*}
has an explicit QTT representation with bond dimension $3$, given by:
\begin{align*}
    D_{\alpha,\beta,\gamma} \! = \!
    \begin{bNiceMatrix}
        \I & \J' & \J
    \end{bNiceMatrix} \! \bowtie \!
    \begin{bNiceMatrix}
        \I & \J' & \J\\
          & \J  &   \\
          &    & \J'  \\
    \end{bNiceMatrix}^{\bowtie (c-2)} \!\!\!\!\bowtie\!
    \begin{bNiceMatrix}
        \alpha \I + \beta \J + \gamma \J'\\
          \gamma \J  \\
          \beta \J' \\
    \end{bNiceMatrix}.
\end{align*}
\end{lemX}
Here, $\bowtie$ denotes the inner core product.

\end{document}